# Prompt-Driven Development with Claude Code: Building a Complete TUI Framework for the Ring Programming Language

Mahmoud Samir Fayed, King Saud University (Riyadh, Saudi Arabia)

Ahmed Samir Fayed, SANS (Jeddah, Saudi Arabia)

## Abstract

Large language models (LLMs) are increasingly used in software development, yet their ability to generate and maintain large, multi-module systems through natural-language interaction remains insufficiently characterized. This study presents an empirical analysis of developing a 7,420-line Terminal User Interface (TUI) framework for the Ring programming language—completed in roughly ten hours of active work spread across three days—using a purely prompt-driven workflow with Claude Code (Opus 4.5). The system was produced through 107 prompts: 21 feature requests, 72 bug-fix prompts, 9 prompts sharing information from Ring documentation, 4 prompts providing architectural guidance, and 1 prompt dedicated to generating documentation. Development progressed across five phases, with the Window Manager phase requiring the most interaction (35 prompts), followed by complex UI systems (25) and controls expansion (20). Bug-related prompts covered redraw issues, event-handling faults, runtime errors, and layout inconsistencies, while feature requests focused primarily on new widgets, window-manager capabilities, and advanced UI components. Most prompts were short, reflecting a highly iterative workflow in which the human role was limited to specifying requirements, validating behavior, and issuing corrective prompts—without writing any code manually. The resulting framework includes a complete windowing subsystem, event-driven architecture, interactive widgets, hierarchical menus, grid and tree components, tab controls, and a multi-window desktop environment. By combining quantitative prompt analysis with qualitative assessment of model behavior, this study provides empirical evidence that modern LLMs can sustain architectural coherence and support the construction of production-grade tooling for emerging programming languages, highlighting prompt-driven development as a viable methodology within software-engineering practice.

**Keywords:** Large Language Models (LLMs), Prompt-Driven Development, Claude Code (Opus 4.5), Human–AI collaboration, Terminal User Interface (TUI), Ring programming language.

# 1. Introduction

Large language models (LLMs) have rapidly transformed the landscape of software development, enabling developers to express intent, describe functionality, and refine systems through natural-language interaction rather than traditional coding. These capabilities have sparked growing interest in understanding how LLMs can participate in or even lead complex software-engineering tasks. While prior work has explored code generation, automated bug fixing, and conversational programming assistance, the ability of LLMs to design, implement, and maintain cohesive multi-module systems remains insufficiently examined. There is limited empirical evidence on how LLMs behave when asked to construct full-scale frameworks that require architectural consistency, state management, event handling, and long-range reasoning across thousands of lines of code [1-7].

The Ring programming language provides a compelling context for exploring these questions. Ring is a lightweight, flexible, and multi-paradigm language designed to support natural-language programming, scripting, application development, and domain-specific language construction. Its simplicity, readability, and small runtime footprint make it particularly suitable for educational use, rapid prototyping, and tooling for emerging ecosystems. Ring's top-down execution model, dynamic typing, and built-in support for console, GUI, and web programming offer a unique blend of accessibility and expressive power. However, despite its versatility, the Ring ecosystem remains relatively young, with limited native libraries for advanced user interfaces—especially in terminal environments [8-12].

PWCT2 (Programming Without Coding Technology 2) further strengthens this context by providing a visual programming environment that supports Ring as a first-class target language. PWCT2 enables developers to construct applications through hierarchical visual components rather than textual code, making it accessible to beginners while remaining powerful enough for professional development. Its design emphasizes clarity, modularity, and step-based program construction, allowing users to generate Ring code and supports importing Ring code too. Because PWCT2 relies on the richness of the Ring ecosystem, the availability of robust libraries directly enhances its expressive capabilities [13-18].

This paper presents a detailed case study of developing a complete Terminal User Interface (TUI) framework for the Ring programming language using a purely prompt-driven workflow powered by Claude Code (Opus 4.5). Over the course of development, all source code—approximately 7,420 lines—was generated or iteratively refined through natural-language prompts. The human role was limited to specifying requirements, guiding architectural decisions, identifying inconsistencies, and providing error-driven corrections. No manual code was written. This process offers a unique opportunity to observe how an advanced LLM handles system-level design, maintains internal coherence across modules, and responds to iterative refinement in a real-world development scenario.

The resulting framework is substantial: it includes a full windowing system, an event-driven architecture, interactive widgets, data-oriented controls, menu and navigation components, and a multi-window desktop environment—all implemented within the constraints of a terminal interface. Beyond demonstrating the technical feasibility of LLM-generated frameworks, this project highlights the potential of prompt-driven development as a methodology for accelerating ecosystem growth in emerging programming languages. For languages like Ring, which rely on community-driven tooling, LLM-assisted development may offer a practical path toward richer libraries, improved developer experience, and faster adoption. By documenting the development process, analyzing the model's behavior, and evaluating the resulting framework, this study contributes empirical insights into the capabilities and limitations of LLM-assisted software engineering. It also illustrates how prompt-driven workflows can support the creation of production-grade tooling and expands the conversation around human–AI collaboration in programming-language ecosystems.

This study makes the following key contributions:

• It delivers one of the first empirical case studies demonstrating that a large, multi-module software framework—a 7,420-line TUI system—can be developed entirely through prompt-driven interaction with an LLM.

• It provides a quantitative analysis of the development workflow, including prompt categories, iteration patterns, and phase-specific complexity, offering insight into how an LLM maintains architectural coherence over extended sessions.

• It offers a qualitative evaluation of Claude Code's behavior, identifying strengths and limitations in handling state management, event-driven logic, and language-specific constraints within the Ring programming language.

• It contributes a fully functional TUI framework that expands the Ring ecosystem and demonstrates the practical value of LLM-assisted development for emerging programming languages.

The remainder of this paper is organized as follows. Section 2 reviews related work on LLM-assisted software engineering, highlighting gaps in current research and situating this study within the emerging literature on prompt-driven development. Section 3 details the materials and methods, including the development environment, evaluation setup, and preliminary experiments used to assess Claude Code's baseline capabilities with the Ring language. It also presents the prompts used to construct the full TUI framework and describes the evolution of its architecture, components, and demo modules. This section also presents the TUI framework demo. Section 4 analyzes the development workflow through quantitative and qualitative lenses, examining prompt patterns, model behavior, and architectural consistency, and discusses the implications of these findings for human-AI collaboration and tooling in emerging languages. Finally, Section 5 concludes the paper by summarizing key insights and outlining directions for future research.

## 2. Related Work

Research on large language models (LLMs) in software engineering has expanded rapidly, with increasing attention to how these systems support code generation, debugging, and architectural reasoning. Early studies primarily examined small, isolated programming tasks, showing that models could synthesize functions, complete code fragments, and respond to natural-language instructions. While these works established the feasibility of LLM-assisted programming, they did not investigate the challenges involved in constructing large, multi-module software systems [19-27].

More recent research has explored prompt-driven development, in which developers iteratively guide LLMs through natural-language instructions to build entire applications. Findings indicate that LLMs can maintain short-term context and follow high-level design intent, but they frequently struggle with long-range consistency, state management, and cross-module coherence. These limitations become more pronounced as system size and architectural complexity increase [28-34].

Another relevant line of work examines LLM-generated user interfaces, particularly in graphical environments. Prior studies have evaluated models' ability to produce HTML/CSS layouts, mobile UI prototypes, and GUI components. However, research on LLM-generated terminal-based user interfaces remains limited. Text-based UI frameworks require precise control over ASCII rendering, event loops, redraw logic, and input handling—capabilities that have not been extensively evaluated in the context of LLM-driven development [35-37].

A parallel thread investigates LLM support for emerging or programming languages. Existing studies show that LLMs can learn the syntax and idioms of niche languages from limited examples, yet there is little empirical evidence regarding their ability to generate production-grade tooling or frameworks for such languages. This gap is particularly notable for languages with smaller ecosystems, where tooling is often sparse and community-driven [38-40].

Finally, research on LLM adaptability to language-specific constraints—such as execution order, reference semantics, and library conventions—suggests that models often require iterative correction to align with a language's rules. Although several studies document this behavior, there is limited understanding of how these adaptation challenges scale when building large, interconnected systems [41-45].

Taken together, the existing literature leaves several open gaps: the lack of empirical studies on LLM-driven development of large, multi-module frameworks; limited exploration of LLM-generated terminal-based interfaces; and minimal evidence regarding LLM performance in emerging programming languages. These gaps highlight the need for deeper investigation into how modern LLMs behave when tasked with constructing complex, coherent software systems under real-world constraints.

## 3. Materials and Methods

The development of the TUI framework was conducted using **Claude Code (Opus 4.5)** through its web-based interface under a Pro subscription (USD 20). All code generation and refinement occurred within this environment. The complete development process required approximately ten hours, distributed across three separate days [46].

To execute and validate the generated source code, we used the **Ring programming language, version 1.25**. Ring is a modern dynamic programming language that comes with features similar to Lua, Python, Ruby and Visual Basic [47]. All components of the framework—including the windowing system, event manager, widgets, and data controls—were tested directly within the Ring runtime to ensure correctness, stability, and compatibility with the language's standard libraries and terminal capabilities.

### 3.1 Preliminary Evaluation of Claude Code

Before initiating development of the full Terminal User Interface (TUI) framework, we conducted a series of exploratory tasks to evaluate Claude Code's ability to generate correct, maintainable, and idiomatic Ring code. These tasks were designed to probe the model's understanding of Ring's syntax, its responsiveness to natural-language prompts, and its capacity to follow architectural intent.

The tasks included:

- Generating descriptive content about Ring.
- Implementing the Observer design pattern in Ring [48].
- Applying the Model–View–Controller (MVC) design pattern in Ring [49-50].
- Implementing statistical functions (max, min, average) with RingQt [9,51].
- Creating a simple ToDo application using RingQt [9,52].

Each coding task was completed using prompt-driven interaction only, without manual code editing. We evaluated the outputs for correctness, clarity, and alignment with Ring's conventions. These experiments demonstrated that Claude Code could reliably generate functional Ring programs, respond to iterative refinements, and maintain conceptual consistency across prompts. Based on these results, we proceeded with confidence to develop the full TUI framework using the same methodology.

The preliminary evaluation tasks were conducted using Claude Code (Sonnet 4.5) through its free-tier subscription. These tasks included generating simple applications, implementing design patterns, and producing descriptive content to assess the model's baseline capabilities with the Ring programming language. In contrast, the full development of the Terminal User Interface (TUI) framework was carried out using Claude Code (Opus 4.5) under a paid subscription.

### 3.1.1 Generating descriptive content about Ring

To evaluate Claude Code's ability to generate contextual, non-code output, we prompted the model to describe the creator of the Ring programming language and explain what makes Ring's design distinctive. The model responded with a coherent and informative summary, identifying Mahmoud Samir Fayed as the creator, noting the language's development timeline, and referencing his academic background and early programming experience. It also highlighted his contributions to both Ring and PWCT, positioning him as a skilled and visionary developer.

Also, the model listed five distinctive features of Ring's design: support for natural-language programming, a visual-programming foundation via PWCT, multi-paradigm flexibility, syntax customization for DSL creation, and practical cross-platform support. These responses demonstrated Claude Code's ability to synthesize biographical and technical information.

### 3.1.2 Implementing the Observer design pattern in Ring

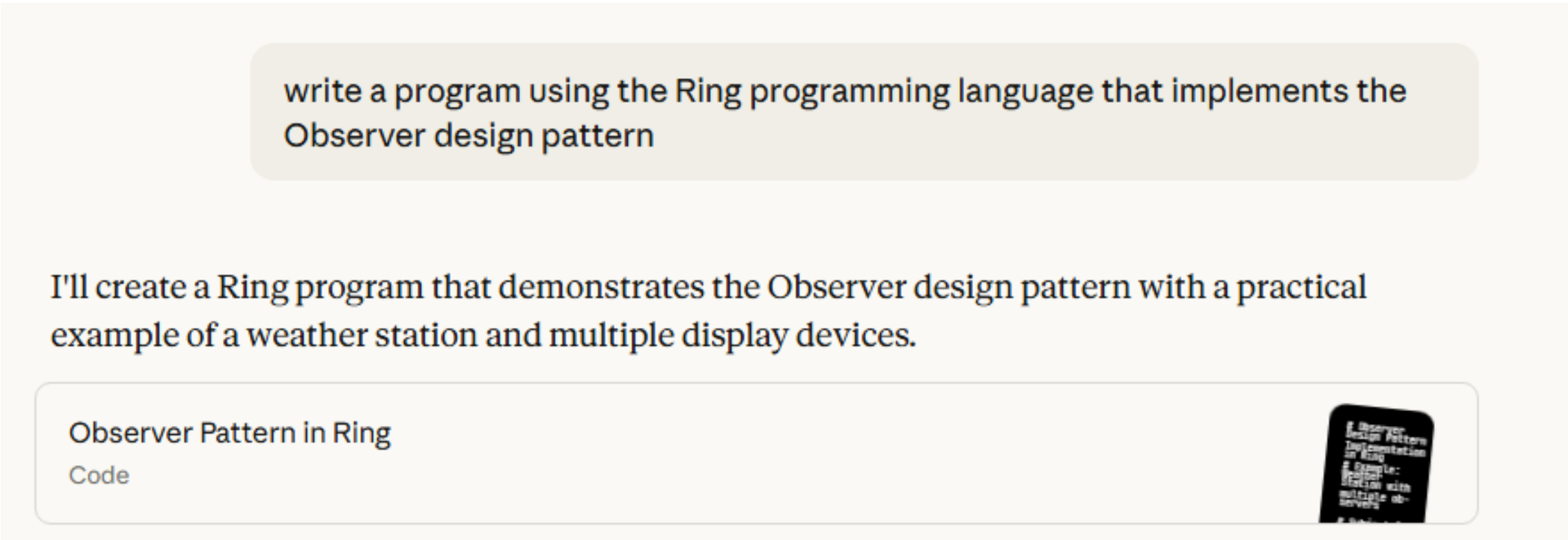

Figure 1. Using Claude Code to implement the Observer Pattern in Ring.

To evaluate Claude Code's ability to implement established design patterns in Ring, we prompted the model to write a complete program demonstrating the Observer pattern (As in figure 1). The model responded with a structured implementation centered around a weather station and multiple display devices, using Ring's object-oriented features to model subjects and observers. The initial code was functional and conceptually correct, but it required reorganization to align with Ring's top-down execution model—where the main program must precede class definitions. After prompting the model to restructure the code accordingly, it produced revised versions that placed the main logic first, followed by class declarations as shown in figure 2. This interaction demonstrated Claude Code's responsiveness to language-specific constraints and its ability to adapt architectural patterns to Ring's execution semantics. The task confirmed the model's capacity to translate abstract design principles into idiomatic Ring code and highlighted its usefulness for pattern-driven development.

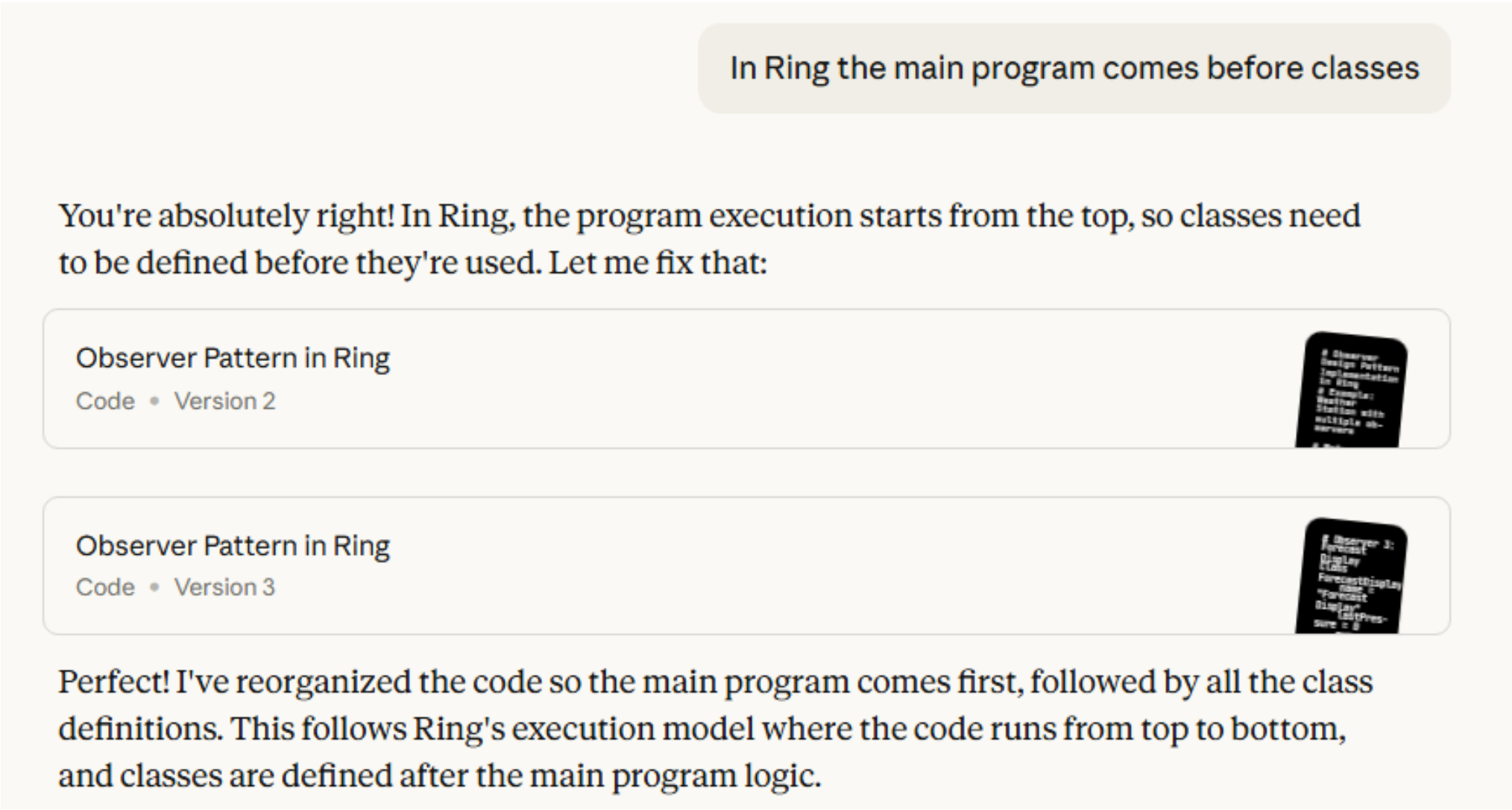

Figure 2. Fixing a problem in the order of instructions.

### 3.1.3 Applying the MVC design pattern

To evaluate Claude Code's ability to implement structured architectural patterns in Ring, we prompted the model to create an example of the Model–View–Controller (MVC) design pattern as shown in figure 3. The model responded with a well-organized implementation of a task management application, dividing responsibilities across model, view, and controller components. The initial output demonstrated a clear separation of concerns: the model handled task data, the view rendered output, and the controller coordinated user interaction.

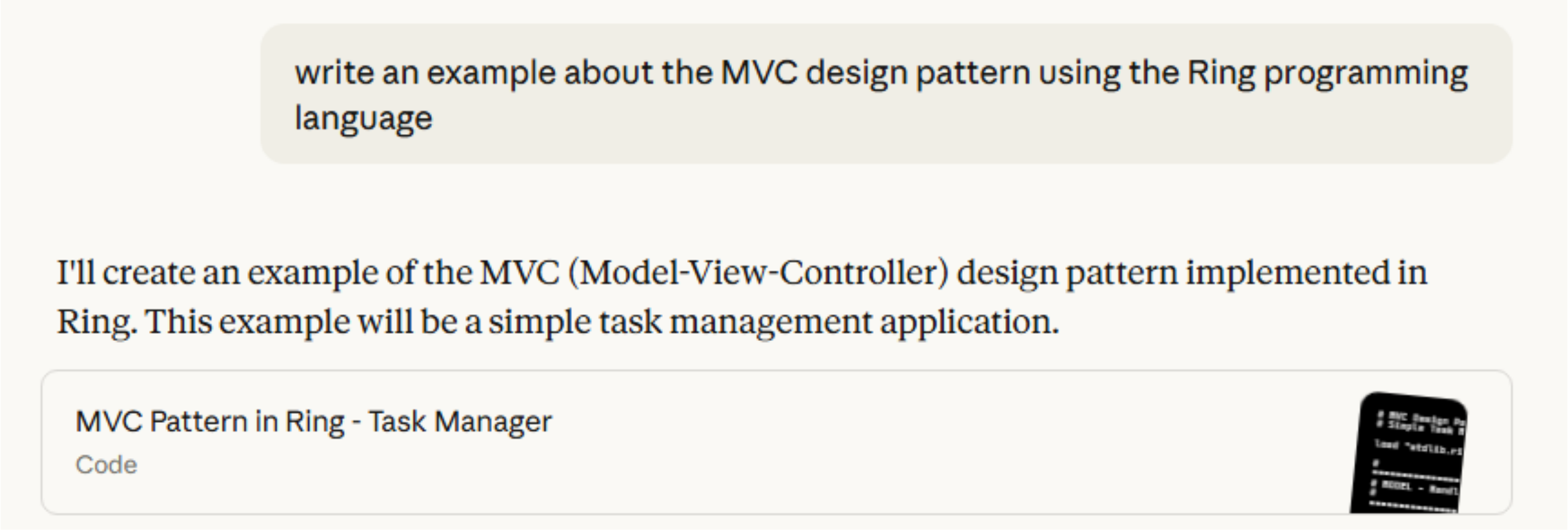

Figure 3. Generating MVC Pattern example.

However, the first version did not fully adhere to Ring's top-down execution model, where the main program must precede class definitions. After prompting the model to reorganize the code, it produced a revised version that placed the main logic first, followed by class declarations. This correction was handled smoothly, and the updated code was both functional and idiomatic.

In subsequent iterations, we identified subtle issues in the toggleComplete() and updateTask() methods. Due to Ring's value-copy behavior with the assignment operator, modifications to list items were not reflected in the original data structure.

We explained that Ring requires the use of the Ref() function to obtain a reference to the actual list item. The model responded with corrected versions that used Ref() appropriately, ensuring that changes to task completion status and task updates were applied directly to the original list.

This task confirmed Claude Code's ability to translate abstract architectural patterns into executable Ring code, adapt to language-specific constraints, and respond to nuanced runtime behaviors through iterative refinement. It also demonstrated the model's capacity to support framework-level development and educational use cases within the Ring ecosystem.

### 3.1.4 Implementing statistical functions with RingQt

To evaluate Claude Code's ability to generate interactive GUI applications using RingQt, we prompted the model to create a program that asks the user to enter five numbers and then displays the maximum, minimum, and average values. The model responded with a functional implementation titled Number Statistics Calculator, using RingQt widgets to collect input and display results. The generated code correctly handled numeric input, performed the required calculations, and presented the output in a user-friendly format. This task demonstrated Claude Code's fluency with RingQt's event-driven model and its ability to integrate basic statistical logic into a graphical interface as shown in figure 4 and figure 5.

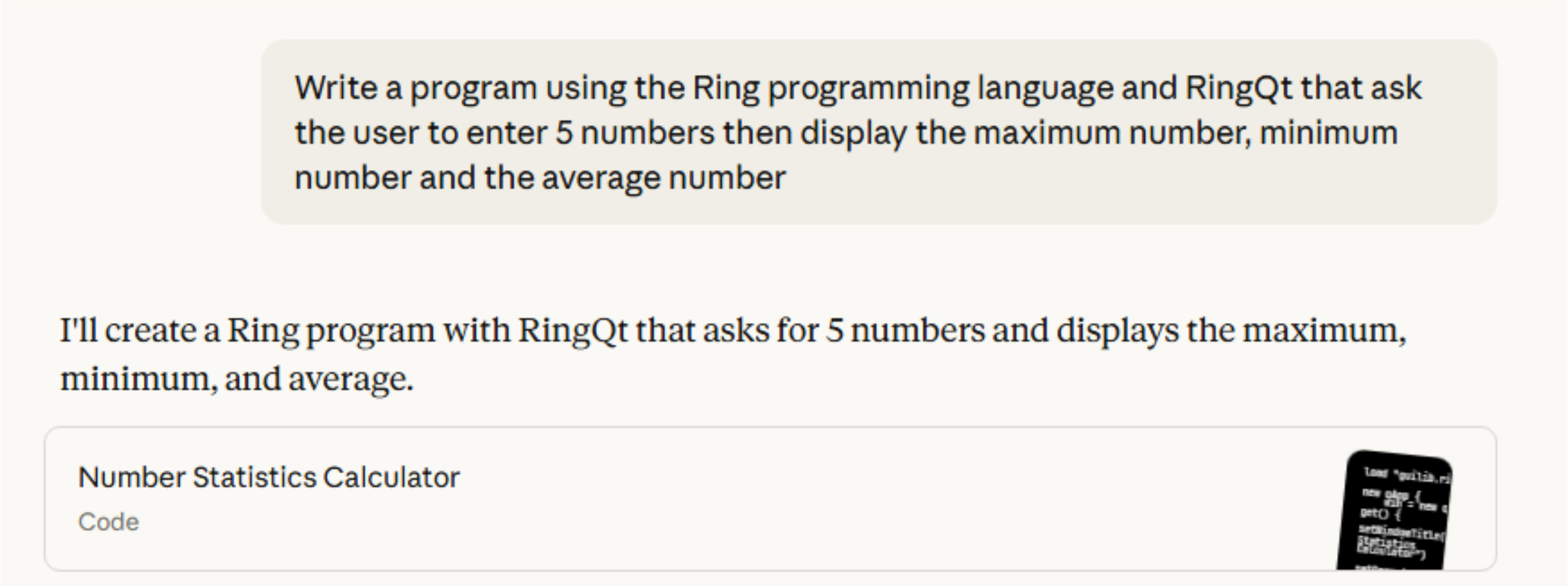

Figure 4. Generating RingQt example.

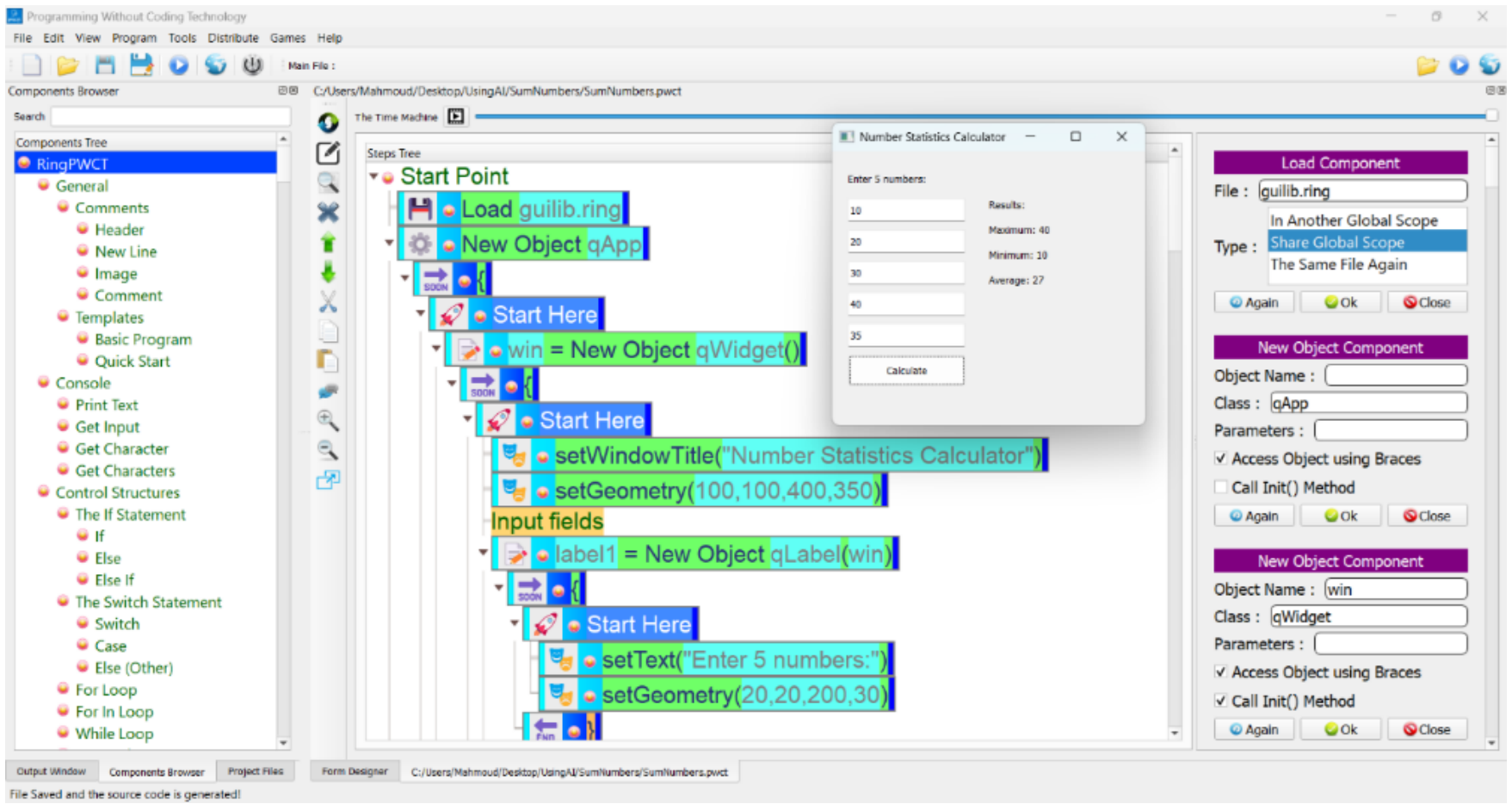

Figure 5. Using the PWCT2 visual programming language to open Ring code.

### 3.1.5 Creating a simple ToDo application using RingQt

The objective of this task was to evaluate Claude Code's ability to develop a complete graphical ToDo application using the Ring programming language and the RingQt framework. Our task was to guide the model through the implementation of a functional application that supports adding, editing, deleting, and viewing tasks. Throughout the process, we observed the model's behavior, identified its mistakes, and documented how it responded to corrections. The interaction with Claude Code unfolded as an iterative development cycle. We provided feedback based on Ring's architecture, syntax, and runtime behavior, and Claude produced updated versions of the code. Each version addressed specific issues, gradually improving the application. This process allowed us to assess the model's capacity to adapt to a language's rules, correct its own errors, and refine its output through repeated guidance.

List of mistakes made by Claude Code:

1. Used the wrong number of parameters with addItem in QListWidget.

2. Used an incorrect event name, setClickedEvent instead of setClickEvent.

3. Failed to use Method for dynamic method resolution when required.

4. Incorrectly configured the Objects Library, leading to runtime errors.

5. Left the view object uninitialized.

6. Introduced an unnecessary CreateView method.

7. Added an unnecessary init function in the view class instead of placing widget creation directly after the class declaration.

8. Created a naming conflict by using the same identifier for both a dialog object and its result value, causing a self-destruction error.

9. Initially treated open_window and openWindow as distinct functions before learning they are equivalent.

By the end of the development process, the ToDo application was fully functional. It supported adding tasks with title, description, and priority; editing existing tasks; deleting tasks with confirmation; and displaying tasks with priority indicators. The final architecture followed a proper controller and view structure using Ring's Objects Library. The generated source code in the Ring language is opened using the PWCT2 visual programming system, where the program appeared as a structured visual tree of components and actions. This demonstrated that the final code could be integrated into a visual programming workflow, further validating the completeness of the implementation. This project illustrates how Claude Code can learn from structured feedback, correct its mistakes, and ultimately produce a complete working application through collaborative refinement.

Figure 6 shows a screenshot of the PWCT2 (Programming Without Coding Technology). At the center is a visual logic tree representing the structure and flow of a ToDo application. Each block corresponds to a command or object in the Ring language, organized hierarchically to reflect program execution. The left panel contains categorized components such as General, Console, Graphics, Web, and Loops, which users can drag into the logic tree. The right panel displays properties for selected components. On the right, a preview of the GUI window titled "ToDo Application" is visible. It includes input fields for task name and description, buttons for adding, editing, and deleting tasks, and a dropdown menu for selecting task priority (Low, Medium, High). This image confirms that the ToDo application is not only functionally complete but also integrated into PWCT2's visual programming workflow, making it accessible to users who prefer graphical development over traditional text-based coding.

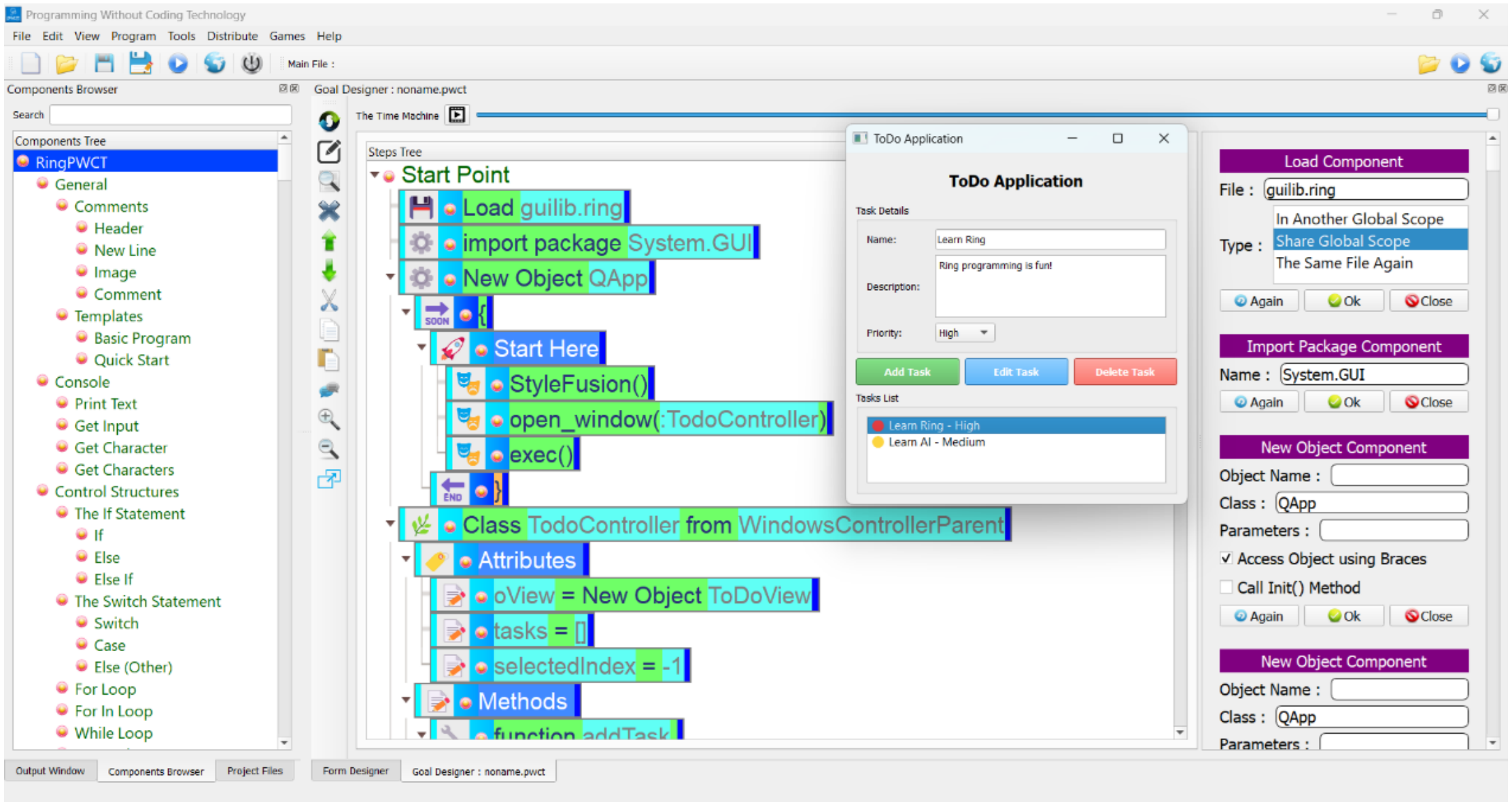

Figure 6. ToDo application in the PWCT2 visual programming language.

## 3.2 Developing the TUI framework

The following table (Table 1) presents the prompts used with Claude Code (Opus 4.5) to create the TUI framework for the Ring programming language. Each prompt is categorized by type: Features, Code Structure, Information from Ring Documentation, Bug Report, etc. The development was completed in roughly ten hours of active work spread across three days.

Table 1. English-language prompts used to construct the TUI framework.

| Index | Prompt | Type |
|---|---|---|
| 1 | Using RogueUtil library in the Ring programming language (Using RogueUtil — Ring 1.25.0 documentation) - start developing a TUI library written in Ring language, the programming paradigm is object oriented, and I expect it to be event-driven and accept input from the keyboard or the mouse at the same time. At first start with a class called Kernel that abstract RogueUtil functions. Then create class called events that contains the main events loop, enable registering events and objects. Create classes for different widgets like (Windows, Label, TextBox, EditBox, ListBox, Combobox, Checkbox) - Then create a demo program that uses these classes - This should work in the terminal | Features |
| 2 | In Ring the main function comes first | Code Structure |
| 3 | These are the functions provided by RougeUtil library: void locate(int x, int y) int getch(void) int kbhit(void) void gotoxy(int x, int y) int getkey(void) int nb_getch(void) char *getANSIColor(const int c) char *getANSIBgColor(const int c) void setColor(int c) void setBackgroundColor(int c) int saveDefaultColor(void) void resetColor(void) void cls(void) void setString(char *str) void setChar(char ch) void setCursorVisibility(char visible) void hidecursor(void) void showcursor(void) void msleep(unsigned int ms) int trows(void) int tcols(void) void anykey(char *msg) void setConsoleTitle(char *title) char *getUsername(void) void printXY(int x, int y, char *msg) void echoon(void) void echooff(void) List *getmouseinfo(void) void enablemouse(void) void disablemouse(void) | Information from Ring documentation |
| 4 | I get syntax error at this line: call handler[2](event) handler[2] must be stored in specific variable first | Bug report (Syntax Error) |
| 5 | I get this error: Line 199 Error (R4) : Stack Overflow In method echooff() in file demo.ring Called from line 199 In method echooff() in file demo.ring Called from line 132 In method init() in file demo.ring Called from line 13 In function main() in file demo.ring Called from line 11 in file demo.ring ---{ Recursion Depth: 996 }--- | Bug report (Runtime Error) |
| 6 | I get this error: Line 142 Error (R4) : Stack Overflow In method gotoxy() in file demo.ring Called from line 142 In method gotoxy() in file demo.ring Called from line 220 In method drawrect() in file demo.ring Called from line 438 In method draw() in file demo.ring Called from line 21 In function main() in file demo.ring Called from line 11 in file demo.ring ---{ Recursion Depth: 995 }--- | Bug report (Runtime Error) |
| 7 | Update this TUI library written in Ring programming language and add the button class: | Features |
| 8 | Add Editbox class (editing multiple lines text) | Features |
| 9 | I get this error: Line 638 Error (R3) : Calling Function without definition: split In method settext() in file C:/Users/Mahmoud/Desktop/UsingAI/tui/demo.ring | Bug report (Runtime Error) |
| 10 | Ring comes with split() function in stdlibcore.ring | Information from Ring documentation |
| 11 | Update the listbox class to keep the selected item when moving the focus | Features |
| 12 | Update the Editbox class to enable moving between lines using Enter key. Enable unlimited number of lines | Features |
| 13 | Update the library so I can move the focus between controls using the mouse | Features |
| 14 | I get this error Line 249 Error (R4) : Stack Overflow In method getmouseinfo() in file C:/Users/Mahmoud/Desktop/UsingAI/tui/demo.ring Called from line 249 In method getmouseinfo() in file C:/Users/Mahmoud/Desktop/UsingAI/tui/demo.ring Called from line 413 In method run() in file C:/Users/Mahmoud/Desktop/UsingAI/tui/demo.ring Called from line 153 In function main() in file C:/Users/Mahmoud/Desktop/UsingAI/tui/demo.ring Called from line 12 in file C:/Users/Mahmoud/Desktop/UsingAI/tui/demo.ring ---{ Recursion Depth: 996 }--- | Bug report (Runtime Error) |
| 15 | I don't know why the mouse doesn't work; this is an example about using the mouse in RogueUtil library | Information from Ring documentation |
| 16 | Now the mouse works to move the focus between controls. Update the library so the mouse can work too at each control level. For example, in Textbox and editbox classes it can move the cursor at specific position. In listbox class it can select specific item. In checkbox class it can invert the checked state. do what you find suitable in a good TUI library | Features |
| 17 | Great, but I discovered a bug, when I click on an item in the listbox, the previous item is selected instead of the clicked item | Bug report |
| 18 | The problem stays. Note (this problem doesn't happen in editbox class) | Bug report |
| 19 | The problem is not fixed; I will check it later. For now, add a class to draw rectangles (single line, double lines, filled or not filled) | Features |
| 20 | it doesn't look good it looks like this | Bug report + Screen shot |
| 21 | IT LOOKS LIKE THIS NOW | Bug report + screen shot |
| 22 | The listbox problem is fixed | Information |
| 23 | Add a table/Grid class that contains multiple columns where I can move between rows or columns using arrows. And I have edit option if enable when I press enter on an item I can edit it's text like textbox. columns could have headers (optional) and rows could have (row number - optional) - i.e. a professional Grid class so i can use it in the future to edit data of my database datable. just focus on the UI part only | Features |
| 24 | I get this error: Line 24 Error (R24) : Using uninitialized variable: kernel In function main() in file C:/Users/Mahmoud/Desktop/UsingAI/tui/demo.ring Called from line 12 in file C:/Users/Mahmoud/Desktop/UsingAI/tui/demo.ring | Bug report (Runtime Error) |
| 25 | Update the latest version of the library so when using a Grid class, we have a space between the line number and the first column (like the ID) also enable modifying a cell in the Grid class when we press enter at any cell: | Features |
| 26 | replace KEY_PGDN with KEY_PGDOWN - These are the defined constants by RoguteUtil library in ring language | Information from Ring documentation |
| 27 | error message: Line 1509 Error (R24) : Using uninitialized variable: key_pgdn In method handlenavigationmode() in file | Bug report (runtime error) |
| 28 | In the Grid cell when I press Enter nothing happens, I can't edit the cell | Bug report |
| 29 | When I press enter to edit a cell, the content is replaced with strange number, I can't edit anything, keep flickering! | Bug report |
| 30 | continuous flickering when trying to modify a cell, suggestion, have a specific loop for the cell edit process | Features |
| 31 | When clicking a cell to edit I get this error: Line 1459 Error (R3) : Calling Function without definition: startediting In method handlenavigationmode() in file C:/Users/Mahmoud/Desktop/UsingAI/tui/demo.ring | Bug report (runtime error) |
| 32 | When using Grid, select cell and press Enter to modify it, I see strange number in the Cell, the screen keeps flickering and can't modify the cell content. Fix this bug so I can edit the text content of any cell inside the gird when using the Grid class: | Bug report |
| 33 | When I press ENTER at the grid cell to edit it, nothing happens! | Bug report |
| 34 | Now it works. When editing a cell enable using the mouse to move the cursor, also if the mouse clicks away end the edit process and send the click to the main event loop to process it maybe it moves the focus to other control | Features |
| 35 | Update the library to add Support for Menu bar where we could have menus like (File Edit Help) and clicking on a menu open (sub menu) and and menus could be nested (i.e. clicking on an item could open another submenu) - Support using mouse/keyboard to move between items. Items could be normal items or checked items (true/false) - we can have a line separator between items. I.e. create professional implementation classes that provide good features | Features |
| 36 | When I click file or any menu this is what i GET, i don't see the menu i see just two strange characters, suggestion - Use separate loop when a menu is activated as we do when edit a cell inside Grid | Bug report + Screen Shot |
| 37 | I still don't see the menu items when clicking on File | Bug report |
| 38 | I still don't see the items, this is what I GET - Suggestions - Learn from the Grid class and how it display the items | Bug report + screen shot |
| 39 | This is what I get when I click file or other option just ++ | Bug report + screen shot |
| 40 | The same problem. In Ring we use Ref() to get a reference to a list/object because assignment/return copy/return variables by value (not reference) - Documentation: Using References — Ring 1.25.0 documentation | Information from Ring documentation |
| 41 | Now I see the menu, but using arrows doesn't change the active item also clicking on another menu doesn't clear the previous menu: | Bug report + screen shot |
| 42 | When I click on a menu like (File) the sub menu appears for a moment then disappear I can't see it to select an item | Bug report |
| 43 | the problem still exist, the menu appears for a moment (looks great by the way) then disappear | Bug report |
| 44 | Now the problem is fixed, I have another problem when clicking on a checkbox in the menu I get an error message as in the image | Bug report + screen shot |
| 45 | In Ring we have this feature: Callable Functions as Methods Using the Call command we can call anonymous functions Syntax (1):<br>Call Variable([Parameters])<br>Also, we can call the function as a method belongs to the current object Syntax (2):<br>Call { Variable([Parameters]) } | Information from Ring documentation |
| 46 | I get this error: Line 627 Error (R24) : Using uninitialized variable: $statusbar In function drawmenudemoscreen() in file C:/Users/Mahmoud/Desktop/UsingAI/tui/demo.ring | Bug report (runtime error) |
| 47 | Now the menu works, In the Grid demo I noticed strange characters at the left - maybe wrong drawing of lines - Fix these things: | Bug report + screen shot |
| 48 | Wonderful, the demo looks more beautiful now, but what do you think about the grid demo | Information + Screen shot |
| 49 | the city column go outside the grid size: | Bug report + screen shot |
| 50 | when opening the demo and selecting one of the three items to open say the form/grid/menu demo - I want pressing escape return to the main menu of the three demos options. Also add and option to exit the app | Features |
| 51 | now add more classes to the library to have movable/resizable windows that can contains child controls and can be moved and resized. add button to maximum, minimize and close windows. Add a specific demo to these features in the main menu. | Features |
| 52 | Nice start but resizing the window hide its contents. Also switching from one to another leads to crash - remember ref() usage in Ring language | Bug report |
| 53 | This is what I get (drawing is not correctly) - also clicking on a window gives flickering, in Ring we can add lists/objects to lists using add() and ref() to add by reference, then find can find object by reference. Also use the trick of click delay if it's necessary | Bug report + screen shot |
| 54 | Very nice but resizing the window and decreasing its size keep the labels visible (outside the window borders) - Also, can the demo contains more controls inside windows (not just labels) we have many controls in the library | Features |
| 55 | I get this error: Line 735 Error (R20) : Calling function with extra number of parameters In function runwindowdemo() in file C:/Users/Mahmoud/Desktop/UsingAI/tui/demo.ring Called from line 72 In function main() in file C:/Users/Mahmoud/Desktop/UsingAI/tui/demo.ring Called from line 15 in file C:/Users/Mahmoud/Desktop/UsingAI/tui/demo.ring | Bug report (Runtime Error) |
| 56 | Three problems (1) In the grid demo, clicking on a cell using the mouse select the next cell in the same column (not the clicked cell). Also, I want double click using mouse starts cell editing as happens when we press Enter (2) In Multiple Windows Demo - resizing window keep large labels visible (goes outside window width) (3) In multiple windows demo - the window that contains textboxes, clicking on any textbox doesn't activate it, i.e. textboxes are not usable. Also, checkboxes need a delay to avoid enable/disable at the same time | Bug report |
| 57 | Multiple problems (1) In the gird demo when I double click a cell or press enter to edit it, the editing textbox appears one line about the cell (i.e. looks like the cell is moved above one line before editing, where two duplicate cells exist). (2) In the form demo clicking on the button produce this error Line 221 Error (R24) : Using uninitialized variable: txtname supported! \| \| In function Form submitted! Name: () in file C:/Users/Mahmoud/Desktop/UsingAI/tui/demo.ring - revise the other buttons too | Bug report (runtime error) |
| 58 | The grid demo problem is not solved see the image | Bug report + screen shot |
| 59 | The combobox need a click delay similar to other controls to avoid changing items quickly | Bug report |
| 60 | Update the library, add more controls (Tree Control, Tabs Control) - Update the demo and add two demos (Tree Demo, Tabs Demo) - - In the tree control support expanding/unexpanding items by clicking +/- button before each item if the item contains children. Use delay when processing click events as we did with other controls. Support Keyboard/Mouse. | Features |
| 61 | In the tree demo I don't see nested items where I can click parent nodes to show/hide children | Bug report + screen shot |
| 62 | Now the tree control is great, with respect to the tab control demo looks like a mess as in the image | Bug report + screen shot |
| 63 | Too much better but I still see strange text in the background in the Tabs demo | Bug report + screen shot |
| 64 | We still have the problem, looks like this now: | Bug report + screen shot |
| 65 | When we draw windows and move them, we notice flickering. What about adding a buffering system to the library. Each draw operation go to a buffer, like array that store the character, color, position, etc. and once draw operations are done in the buffer, we reflect these operations on screen at once to avoid flickering. if this solution is ok, do it, if you know better solution do it. just avoid flickering | Features |
| 66 | add width method to kernel class so we can have strings with fixed width - Use it in form demo, menu bar demo so displaying messages override old messages and avoid interference | Features |
| 67 | In the Form demo when I click reset button, looks like draw() method needs to be called so I can see the empty controls (i.e. clear values) | Bug report |
| 68 | In the Grid control, selecting a cell draw all of the other cells and leads to flickering, solve this by just redrawing the old cell (that has the focus) and the new cell that will get the focus | Bug report |
| 69 | This is better no flickering when moving between cells using the keyboard, but I see flickering when moving between cells using the mouse | Bug report |
| 70 | Still clicking on a cell using mouse led to flickering. Also, I noticed another problem, changing rows using keyboard doesn't highlight the row number of the current row, and selecting another row using mouse highlight the next row number instead of current selected row | Bug report |
| 71 | The highlighted row problem is fixed. But still see flickering when selecting a cell using mouse also noticed that when selecting a cell using mouse it's opened for edit mode directly, I want this happens when double click, i.e. mouse click only select the cell and don't edit it. Another problem, the line after grid columns headers. have something like \|\| under city cell, i.e. we have extra \| that should be dash - | Bug report |
| 72 | The problem of drawing the line after gird header exist | Bug report + screen shot |
| 73 | It looks like this now | Bug report + screen shot |
| 74 | Now it looks correct, Time to solve the mouse click problem, use delay trick, when we click a cell just select it don't start editing, and avoid flickering as we do with keyboard movement. if we click again on a selected cell then start editing | Bug report |
| 75 | Update the library and add more controls (progress bar, spinner where we increment/decrement a number using buttons and can enter only numbers, vertical/horizonal scrollbars) - Add new demo to the main menu for these controls | Features |
| 76 | It looks like this, I see a lot of ? characters | Bug report + screen shot |
| 77 | It looks very nice now, but sometimes the value of the scroll bar becomes outside range when clicking on it - see image | Bug report + screen shot |
| 78 | the problem of the range in scroll bar still exist, could it be related to text width and overwriting old text | Bug report |
| 79 | In the Grid demo, when we press enter to edit cell, after finishing editing and pressing enter I see flickering where all cells are drawn again, just draw the edited cells. Another problem, In the controls demo which contains progress bars, scrolls bars, we have a vertical scroll bar and a value at the bottom left of it, the value width touches the scroll bar and draw over it, move it 2 spaces to the left to avoid this | Bug report |
| 80 | In the Treeview control, when I use arrows or mouse to change the active node, i see flickering where all nodes are redrawn, just redraw what is necessary (previous node, current node) - I think redraw is required only in specific case where we show/hide node children. | Bug report |
| 81 | Very nice now navigation in the Treeview control using keyboard doesn't cause flickering, but navigation using the mouse does, it redraws also the two Treeview controls in the demo. please fix this. Also add a new demo where we have the (menu bar demo) and the (tabs pages demo) in one screen | Bug report |
| 82 | In the (Menu + Tabs demo) I noticed two problems. The controls in the tabs accept mouse events which is great, but they don't accept keyboard events, i.e. I can use the keyboard when activating a control inside a tab. The second problem when using the menu bar and changing the menu the background becomes blue which means blue rectangles are drawn over the tabs and their controls. Find a way to detect which controls that their place intersects with specific rectangle so when changing a menu those controls only are redrawn. which drawing the screen with a blue background for example do this using a specific control (something like Frame control) which have dimensions two to determine when to redraw such controls. this background could come before other controls, because order matters when redrawing controls. | Bug report |
| 83 | The keyboard problem is fixed, but when changing the menu in the menubar I see black background instead of seeing the Tabs behind the menu and their controls. Maybe you need that the method the determine intersections between controls be in something like the kernel class, and the window manager who have access to all controls can use it. I am not sure do what is right to fix the problem | Bug report |
| 84 | This is too much better, but I have the next problems. In (Menu + Tabs) demo, changing the menu while redrawing controls in background, it keeps a strange line after the menu +-----+ which looks like it's drawn when we display a menu in menu bar. this should be removed. Another problem the (Tabs Demo) which is another option from main menu, no longer support the keyboard events when selecting a control. Another problem (Menubar demo) which is another option from main menu when changing the menu using menubar, the older menu stays in the screen, switching between all menus draws all menus at the same time! - Please fix all of these issues | Bug report |
| 85 | Very nice, now two more issues to fix. The first in the Form Demo, the listbox control suffer from flickering when changing the current item using keyboard or using the mouse, fix the problem as you did for other controls like Grid control or TreeView control. Avoid flickering when using the mouse or the keyboard to change the current item. The second issue in the (Menubar demo) when we change menus, sometimes the nested menu is drawn on the top of label/text in the background, then changing the menu and hiding it doesn't redraw these labels/text. Find a way to redraw them if their place intersects with the menu | Bug report |
| 86 | Very nice, I discovered two more issues, The first when we click on a textbox the cursor doesn't appears until we start typing or press one of the arrow keys, I want the cursor to appear directly once we activate a textbox. The second issue the editbox suffer from flickering when using arrows or clicking on a specific line, don't redraw the editbox if this is not necessary to avoid flickering | Bug report |
| 87 | Very nice the problems are fixed but the textbox problem is fixed only in the form demo, it still exists in (tabs demo) and (menu + tabs demo) where activating a textbox doesn't show the cursor. | Bug report |
| 88 | The problem is not fixed. Leave it for now, I have a more important problem to fix. The menu bar uses a specific events loop in the demo code. Find a way to have this code in the classes itself (like the events manager or whatever suitable class) where when use menubars as we do in form view sample, just add it as a control or using specific method (addMenuBar) where the main events loop check if we have a menubar then call the specific method that handle its keyboard/mouse events. do the same for other demos that uses custom events loop. The point we want high-level library, the user code (demo writer) shouldn't start events loops from scratch but call methods that contains these loops | Bug report |
| 89 | very nice, apply the same idea to other functions like controls demo and treeview demo, both have events loops, we need this to be managed by the events manager as we did for the menubar and tabs | Code Structure |
| 90 | (1) Clicking on a node in the tree produce this error: Line 1043 Error (R24) : Using uninitialized variable: tree in file demo.ring and pressing a key produce this error: Line 1017 Error (R24) : Using uninitialized variable: activetree in file demo.ring - revise methods in events manager maybe you need custom method with custom loop for tree controls (2) Opening controls demo produce this error: Line 1421 Error (R21) : Using operator with values of incorrect type \| \| in file demo.ring (3) In general revise the logic, before previous update things was working but our goal was to use (events manager) instead of custom loops at client code, to fix this think about moving the code that was working before from client code to have its methods in events manager and call such methods when needed, (4) I am just suggesting do the right thing to have good architecture | Bug report (runtime error) |
| 91 | In the window manager demo, moving a window cause huge flickering because all other windows are drawing, also flickering happens when we click on any control in any window. revise drawing logic as we did when we change the current menu in the menubar to draw only controls in the background that needs drawing. | Bug report |
| 92 | Two issues to fix (1) When we maximize a window then minimize it, the maximized window still appears while a new unmaximized window is drawn, fix this while adding a background/shape control to be drawn before all windows. (2) When resizing a window contains listbox and the window size is decreased, the listbox appears outside its parent window. | Bug report |
| 93 | I have the next four issues (1) When we maximized window a window then unmaximize it to restore its original size, in this case draw all windows because during redraw some windows doesn't appear. (2) In the windows demo - user form - clicking on the Email textbox doesn't activate it, I can't use this textbox! (3) I want when I click on a textbox in the form to see the cursor directly (4) Resizing the window then clicking on a textbox or button inside it draws the control outside the window borders! | Bug report |
| 94 | When I resize window then click on textbox or button it's drawn outside window borders as in this image | Bug report + screen shot |
| 95 | Two issues (1) When we resize a window, the checkbox control is drawn outside the window borders (2) When we maximize a window then click to restore its original size, some windows are drawn and the other windows are not drawn, i.e. looks like the window draw window previous windows that comes before it according to creation and don't draw windows that comes after it. (3) In the same windows demo when I click on a textbox I don't see the cursor, just display it when the textbox is activated | Bug report |
| 96 | Too much better many issues are fixed but we have the next issues (1) In the windows demo When we maximize a window then restore it to its original size, the window is hidden, just display all windows in the window system (2) In the tabs demo and in the menubar and tabs demo, clicking on textbox doesn't show the cursor just display it (3) Revise all other controls in the library to be sure that when we add a control to a movable window then resize this window the control can’t be drawn outside the window borders | Bug report |
| 97 | Two problems (1) When maximize a window then return to its original size we have the same problem (not all windows are visible) - trying drawing everything after setting the new size (2) When we click textbox and start typing i see flickering because of redrawing all of other controls in the window | Bug report |
| 98 | The flickering problem is fixed, but the other problem (maximize window then restore size doesn't display all windows) is not fixed. I tried moving all windows away (no intersection) then When I maximize the information window then restore its size, only the (status) window is not drawn and the other two windows (select country) and (user form) are drawn. If I maximized the (status window) then restored, it's size then all windows are drawn correctly. If I maximized (select country window) then restored, it's size then the status window is hidden/not drawn while other windows are drawn. If I maximized (user form) window then restored its size, only two windows appears (user form window) and (information window) while the (select country) and (status) windows are hidden. I hope that this information are useful for you to solve the drawing bug after restoring window size. Another thing to know about, not sure if useful or not, In Ring the assignment operator and Add() function use copy by value for lists/objects, to do this by reference we use the ref() function. | Bug report |
| 99 | These are constants for colors from RogueUtil library, use them for colors instead of using numbers, revise all of the code: BLACK BLUE GREEN CYAN RED MAGENTA BROWN GREY DARKGREY LIGHTBLUE LIGHTGREEN LIGHTCYAN LIGHTRED LIGHTMAGENTA YELLOW WHITE | Information from Ring documentation |
| 100 | (1) In the windows demo I use (Tab key) to move between windows, but this disable using the Tab key to move between controls in the same window, update the library and use another key like (KEY_F12) to move between windows, so I can use the Tab key to move between controls in the same window. (2) I have a similar problem in the Tabs demo where moving between tab pages is done using Tabs, use KEY_F11 for this purpose so we can use Tab to move between controls in the same tab page. (3) Revise the library code/demo to do any required updates related to these two points | Features |
| 101 | Very nice, A little issue, I noticed when switching focus between controls using Tab that the Label control could get the focus! - Such a control should avoid getting the focus and pass it to the next control or the first control if the label is the last one or keep it only if it's the only control in the window. The user uses Tab to move between items that accept an action like typing or clicking, etc. the label control is not one of them. | Bug report |
| 102 | (1) I noticed that the demo code calls the Draw() method for each control alone! update the classes/demo code so calling draw once per window or tab page called be enough to draw controls belongs to the window/tab - Be careful to be sure that this doesn't conflict with other features | Code Structure |
| 103 | Two issues in the menu bar demo (1) click (format menu) then (font menu) then (select size option) here pressing the Right Arrow doesn't open the sub menu where we must press Enter to open it and when we press Enter to open the sub menu, pressing the (Left arrow) doesn't return to the Size option but goes to the (Font option/menu), i.e. two steps backward not just one! (2) When we select (View menu) then (Zoom menu) then press the right arrow to open the sub menu, after this sub menu is opened if we pressed the right arrow again a label from background (Menu bar demo) appears and overwrite the view submenu items. | Bug report |
| 104 | The Tree demo and the Controls demo contain Events Loop, this should be part of the classes as in other demos where we don't need such a complex logic at library user code. Try to do this in a simple way to avoid interference with other features of introducing bugs. maybe using separate methods. Also take in mind Ring anonymous functions doesn't capture variables (not closure) so use global variables in event handlers as in other demos when this is necessary | Code Structure |
| 105 | I get the next errors (1) Line 1074 Error (R24): Using uninitialized variable: $tree1 -----------+ \| in file demo.ring when opening Tree demo (2) when opening controls demo Line 1502 Error (R24) : Using uninitialized variable: $progress1 \| \| in file demo.ring | Bug report (runtime error) |
| 106 | Thanks, development is done, write a complete tutorial for this library to explain in details how to use it step-by-step. Generate the documentation in txt file and follow sphinix documentation style (Sphinix is a python tool for documentation) - Provide enough examples to cover all library features and controls | Documentation |
| 107 | with respect to the latest version of the code (demo.ring) everything works fine but when I open (menubar + tabs) demo then go and open other demos it becomes slower, specially the windows demo and moving windows becomes very slow, this happens only if I opened the (menubar + tabs) demo, if not opened all demos works very fast even if i closed them then opened them many times. | Bug report |

LLM behavior is inherently non-deterministic, and this variability must be acknowledged when interpreting the results of this study. Because large language models evolve across versions and can produce different outputs for the same prompt in separate sessions, the development process described here reflects a specific interaction trajectory rather than a universally reproducible sequence. While the prompts, workflow, and evaluation criteria were consistent, the exact responses generated by Claude Code (Opus 4.5) may differ if the experiment is repeated under future model updates or even within a new session of the same version. This variability does not diminish the validity of the findings; instead, it highlights an important characteristic of prompt-driven development—namely, that outcomes depend not only on prompt design but also on the dynamic behavior of the underlying model. Recognizing this helps situate the results within the broader context of LLM-assisted software engineering and supports a transparent, realistic interpretation of the methodology.

## 3.3 The TUI Framework Demo

To run the demo execute the next commands after installing the Ring programming language version 1.25

- ringpm install tuiframeworkusingclaudecode from ringpackages
- ringpm run tuiframeworkusingclaudecode

In figure 7, below the title, nine numbered options are listed vertically, each corresponding to a specific demo module within the Ring TUI framework. At the bottom, the prompt "Enter choice (1–9):" invites user interaction, this menu serves as the entry point for exploring the modular capabilities of the Ring TUI framework, emphasizing clarity, accessibility, and the expressive power of ASCII-based UI design.

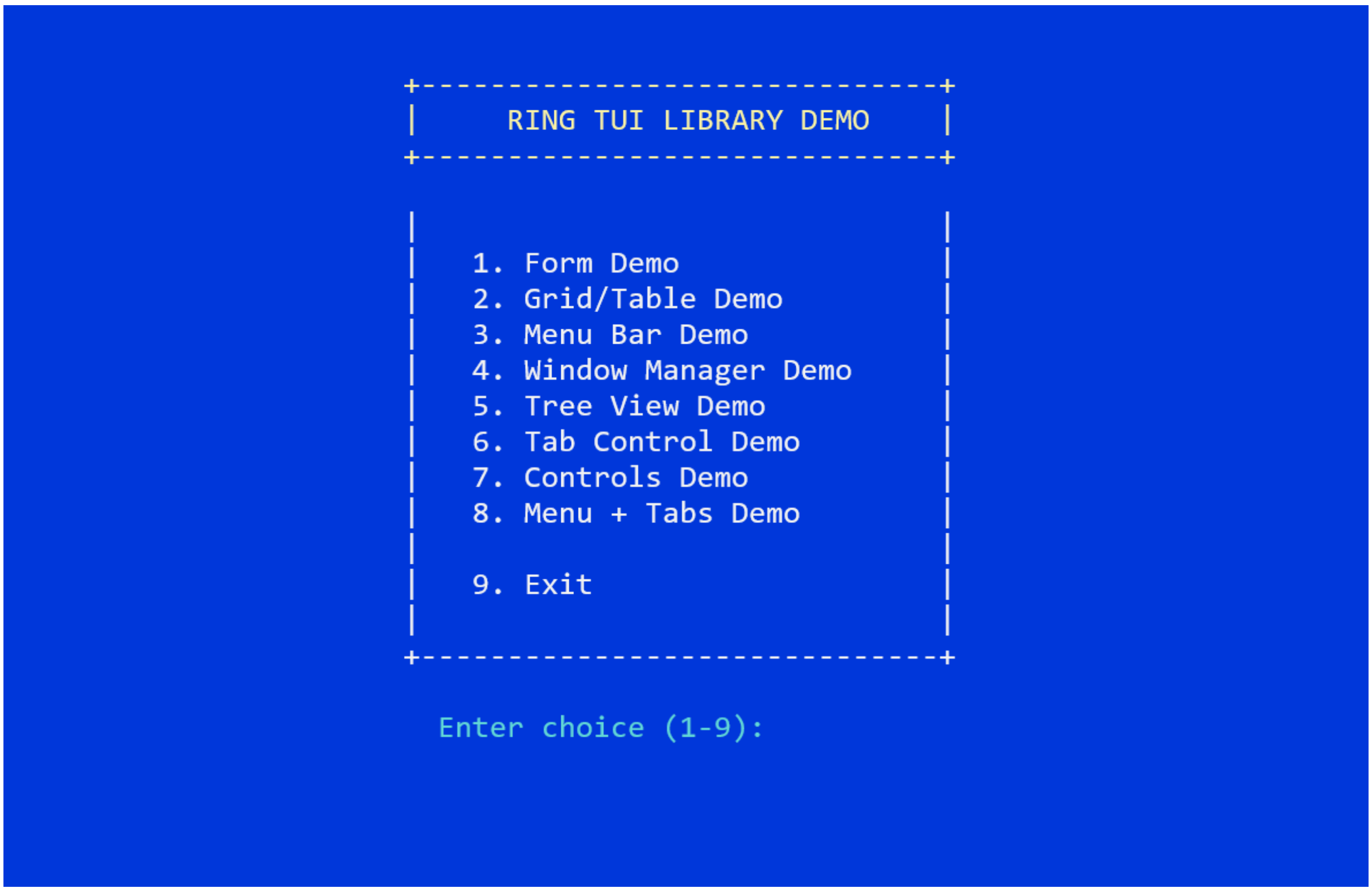

Figure 7. Ring TUI Demo.

Figure 8, illustrates the Form Demo interface of the Ring TUI Library, showcasing a comprehensive text-based user interface rendered entirely within a terminal environment. The layout includes labeled input fields for name and email, interactive checkboxes, a dropdown menu for language selection, and a scrollable country list with highlighted selection. A multiline comment box and action buttons (Submit, Cancel, Reset) simulate GUI-like behavior using ASCII characters, supporting both keyboard and mouse interaction. Instructional cues guide navigation through TAB, arrow keys, and pointer actions, while a status bar reinforces usability. This demo highlights how the Ring TUI framework enables rich, form-driven interaction in non-graphical contexts, supporting accessibility and modular design for terminal applications.

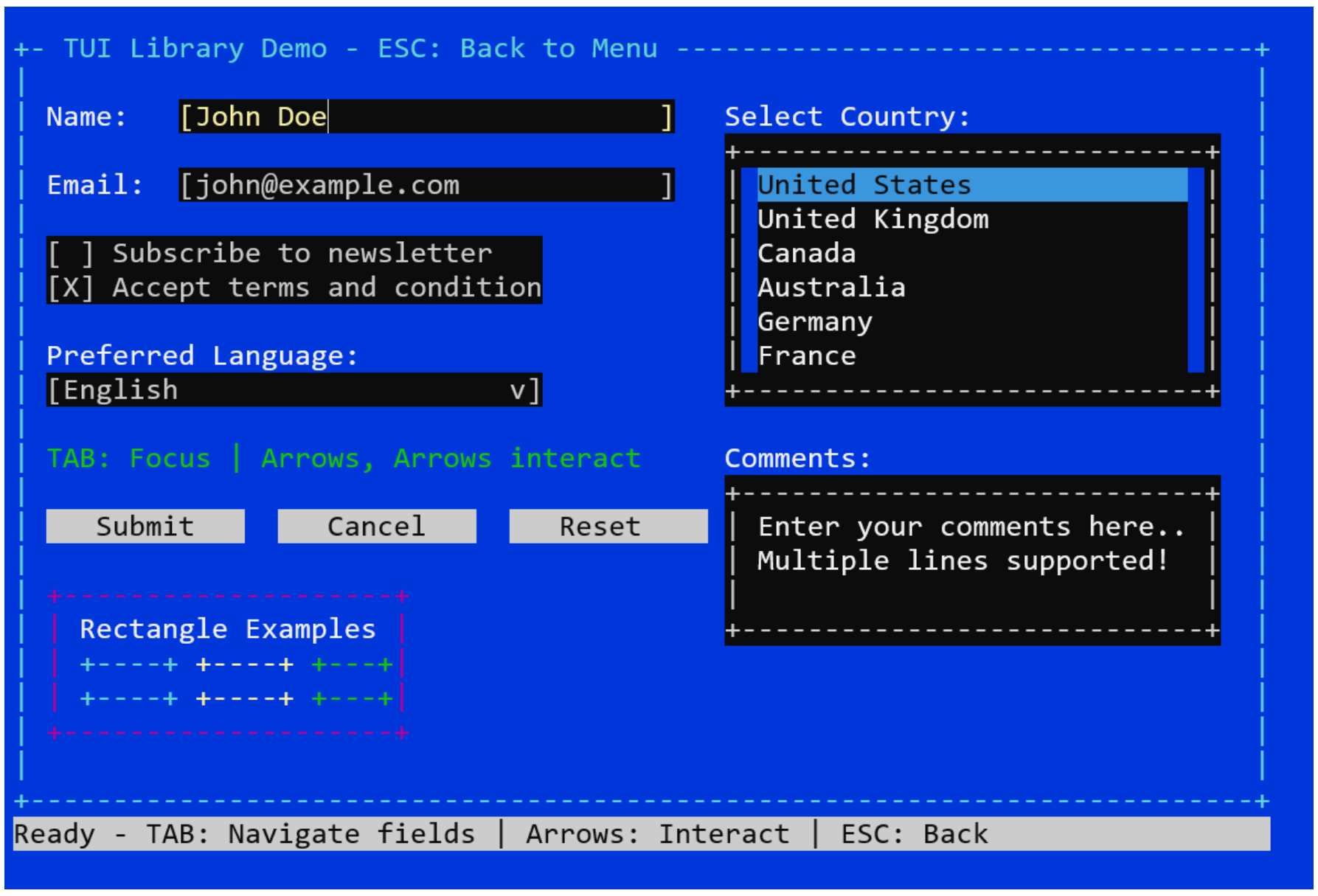

Figure 8. The Form Demo.

Figure 9, presents the Grid/Table Demo interface of the Ring TUI Library, illustrating a structured tabular layout rendered in a terminal using ASCII characters. The table displays ten rows of user data across six columns: serial number (#), ID, Name, Email, Age, and City. Each row represents a distinct record, with realistic sample values to demonstrate data formatting and alignment. Users can navigate the grid using arrow keys or mouse clicks and edit individual cells by pressing ENTER. A footer provides interaction guidance, while the interface maintains visual clarity through consistent column spacing and borders. This demo highlights the framework's capability to support dynamic, spreadsheet-like data presentation and manipulation within a text-based environment, reinforcing its utility for terminal applications requiring structured input and review

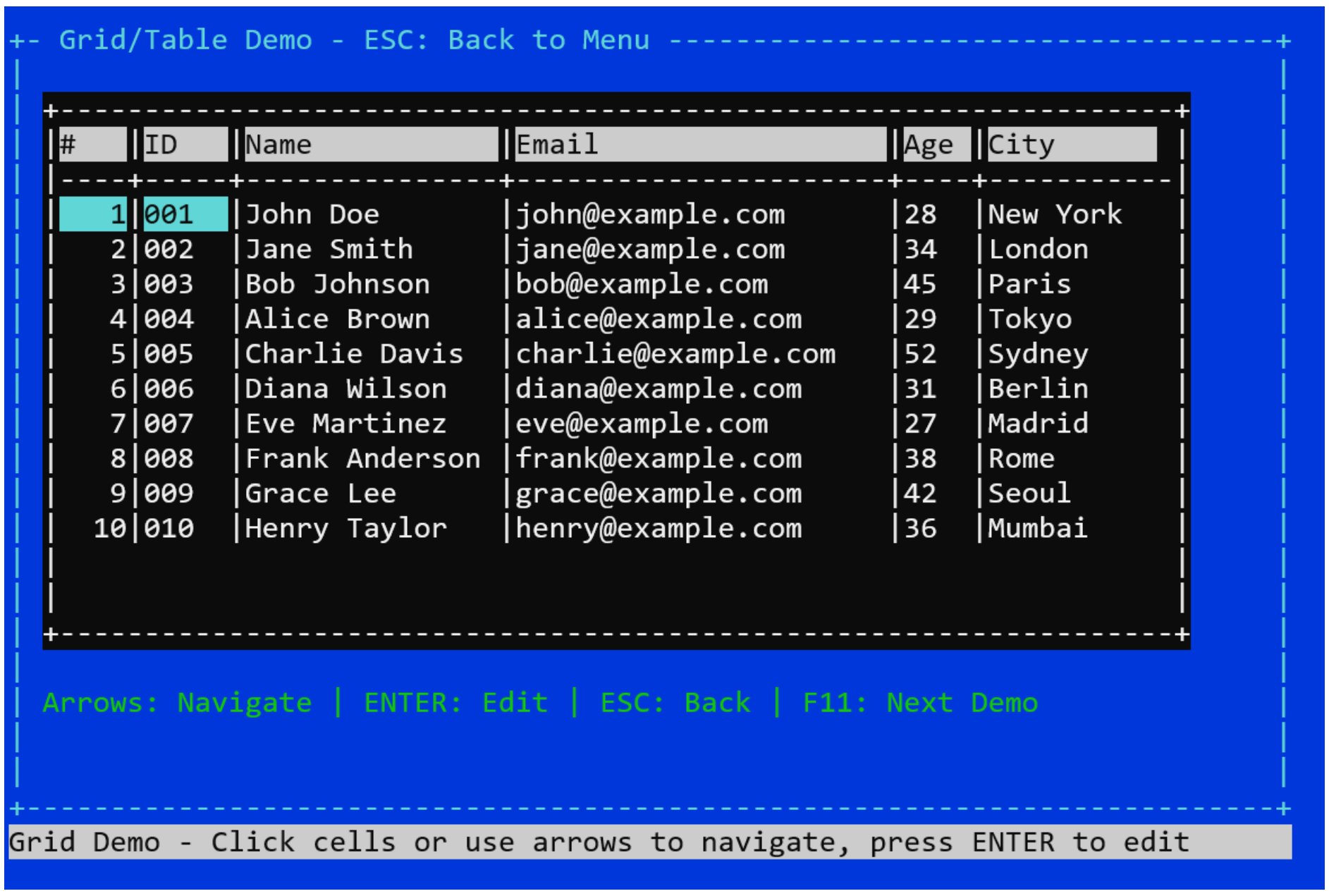

Figure 9. The Grid Demo.

Figure 10, depicts the Menu Bar Demo interface of the Ring TUI Library, demonstrating a fully interactive menu system rendered in a retro, text-based style. The top of the screen features a horizontal menu bar with entries such as File, Edit, View, Format, and Help. The active menu, Format, expands into nested submenus for Font and Size, each containing selectable items. The Font submenu includes typeface options (Arial, Times New Roman, Courier New) and checkbox-style entries for Bold, Italic, and Underline, each with associated keyboard shortcuts (e.g., Ctrl+B). The Size submenu lists font sizes from 10 pt to 18 pt. Below the menu, instructional text guides users to activate menus via F10, navigate with arrow keys, and select or close items using Enter and Escape. This demo highlights the Ring TUI framework's support for hierarchical menus, shortcut keys, checkbox states, and mouse or keyboard interaction, enabling rich navigation and configuration workflows within terminal-based applications.

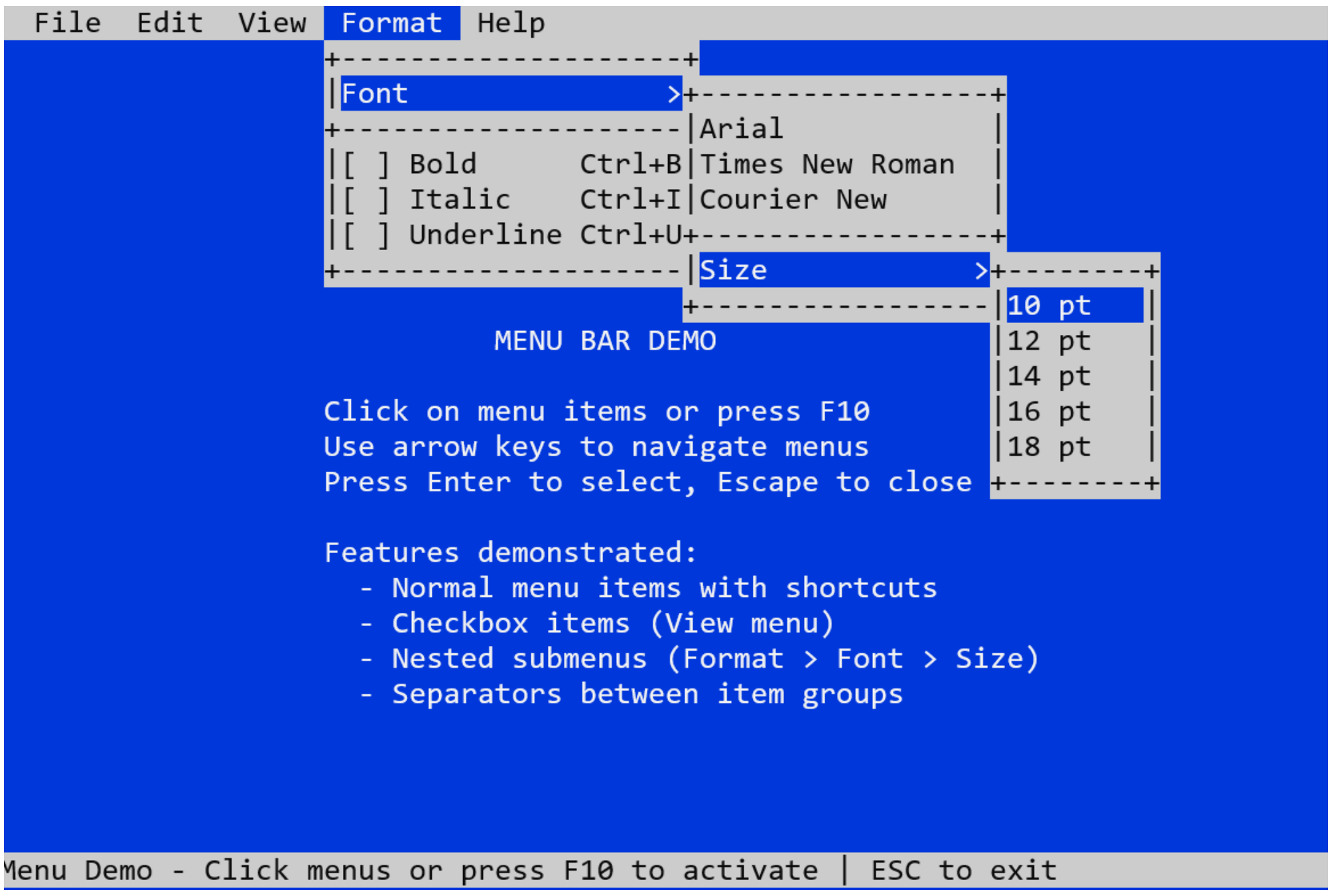

Figure 10. The Menubar demo.

Figure 11, showcases the Window Manager Demo of the Ring TUI Library, simulating a multi-window desktop environment within a text-based terminal interface. The screen features four distinct window panels—Information, User Form, Select Country, and Status—arranged over a blue background and managed through keyboard or mouse interaction. Each window includes a title bar with controls for minimizing, maximizing, and closing, and supports drag-and-resize functionality via designated corners. The Information window outlines available features; the User Form window presents input fields and checkboxes; the Select Country window offers a scrollable list of options; and the Status window displays system metrics such as CPU, memory, disk, and network status. A bottom bar mimics a taskbar, listing active windows and providing quick access. This demo illustrates the Ring TUI framework's capacity to emulate desktop-like behavior in terminal environments, enabling rich multitasking and modular UI composition without graphical dependencies.

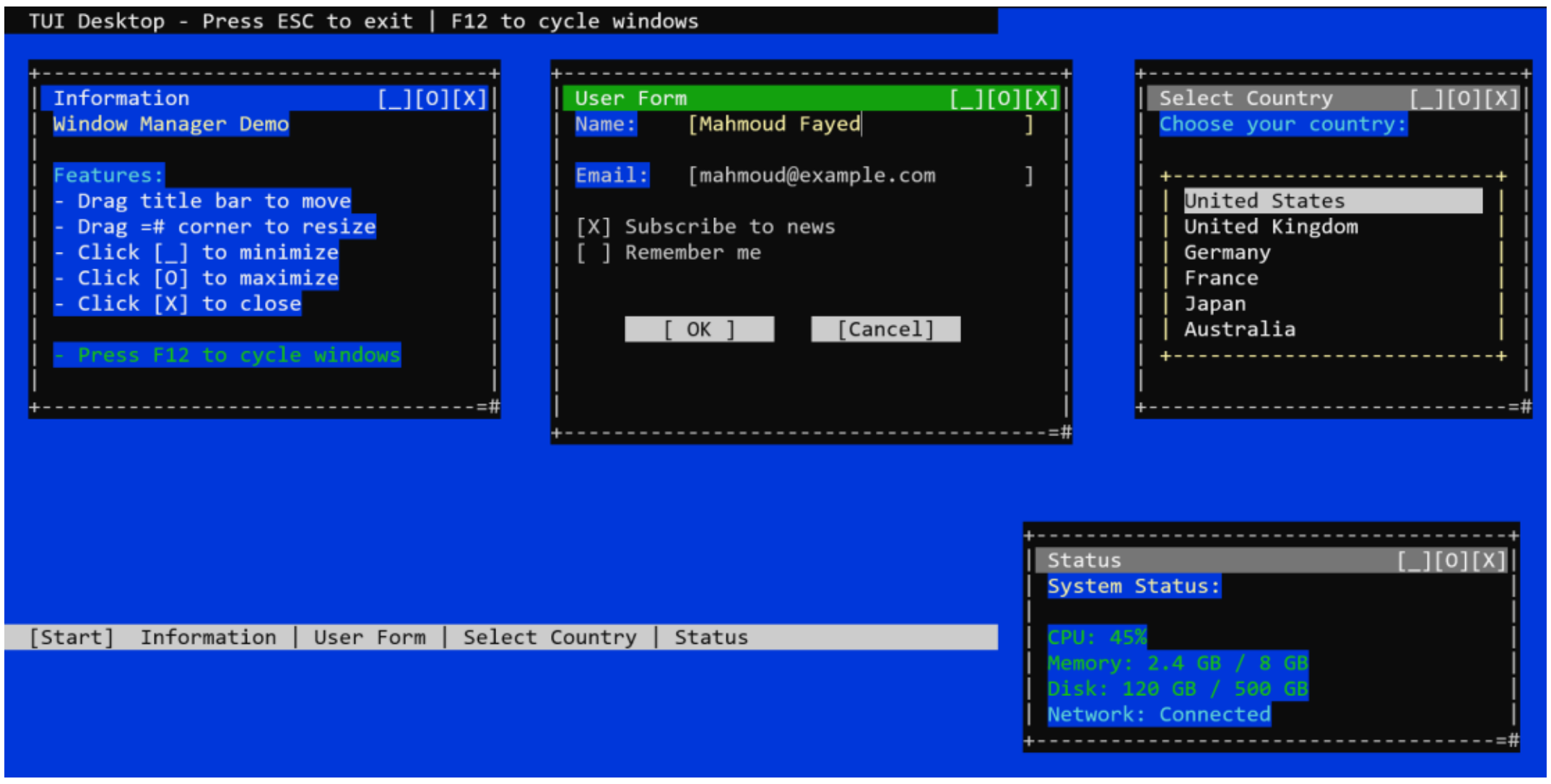

Figure 11. The Windows Demo.

Figure 12, illustrates the Tree View Demo interface of the Ring TUI Library, demonstrating hierarchical data visualization within a text-based terminal environment. The screen is divided into two panels: a file system explorer on the left and an organizational chart on the right. The file system tree begins at the root directory "C:" and expands into nested folders such as Program Files, Microsoft Office, Ring (with executables and library files), Users, and Windows. The organization chart presents a corporate hierarchy starting with the CEO, followed by the CTO and their subordinate managers and roles, as well as the CFO and COO. Navigation instructions at the bottom guide users to expand or collapse nodes using keyboard shortcuts (+/−/Enter/Space) or arrow keys. This demo highlights the Ring TUI framework's ability to represent and interact with complex, nested structures in a terminal interface, supporting both file navigation and abstract data modeling.

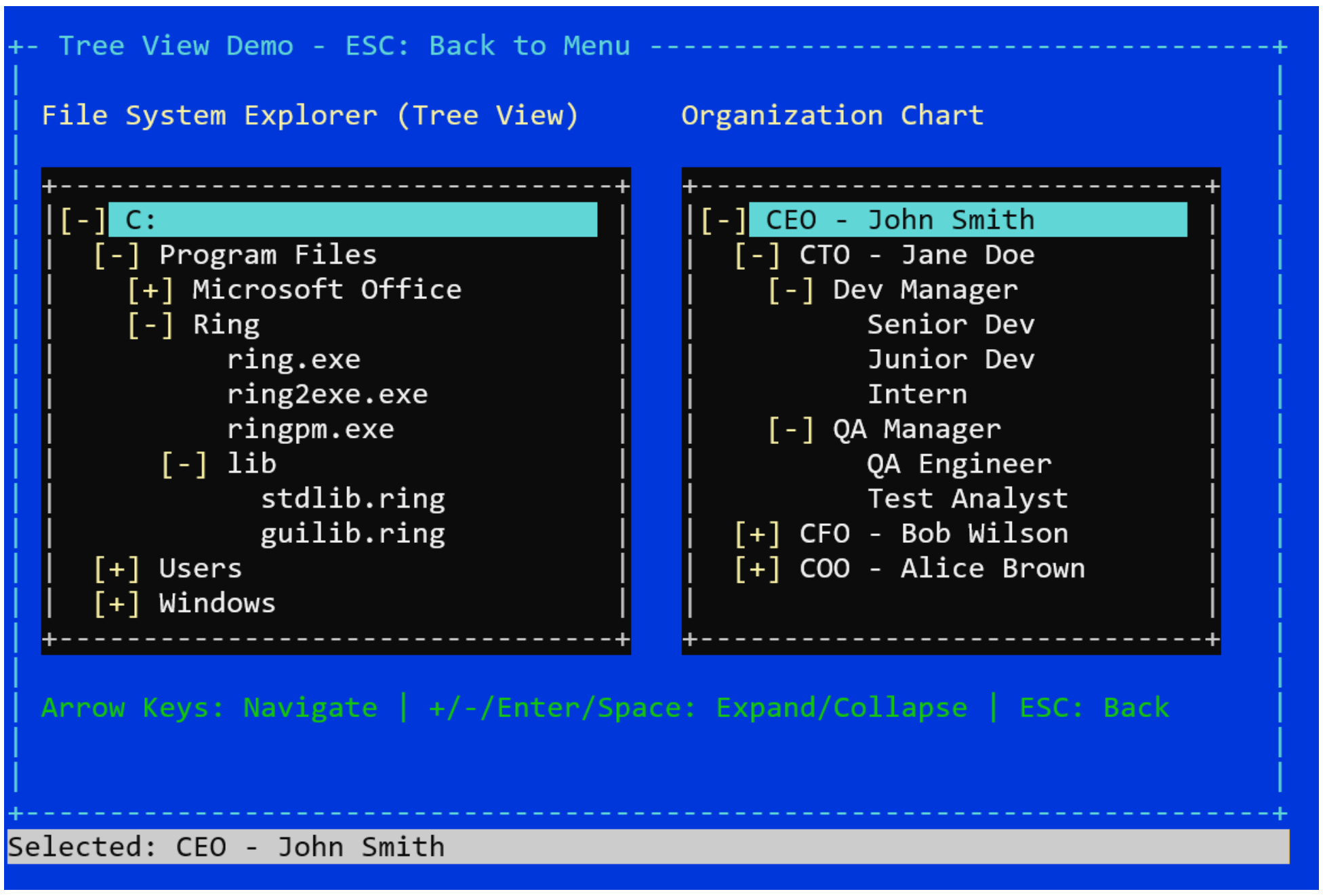

Figure 12. The TreeView Demo.

Figure 13, presents the Tab Control Demo interface of the Ring TUI Library, illustrating a tabbed configuration panel within a text-based terminal environment. The interface features a horizontal tab bar with sections labeled General, Appearance, Network, and About, with the General tab currently active. The content area displays configurable settings including a username field, a language dropdown menu, and two checkbox options—Auto-save enabled and Show notifications—both marked as selected. Navigation instructions at the bottom indicate support for keyboard and mouse interaction, including TAB for control focus, F11 to switch tabs, and ESC to exit. This demo exemplifies the Ring TUI framework's ability to organize interface elements into modular, navigable tabs, enhancing clarity and user experience in terminal-based applications.

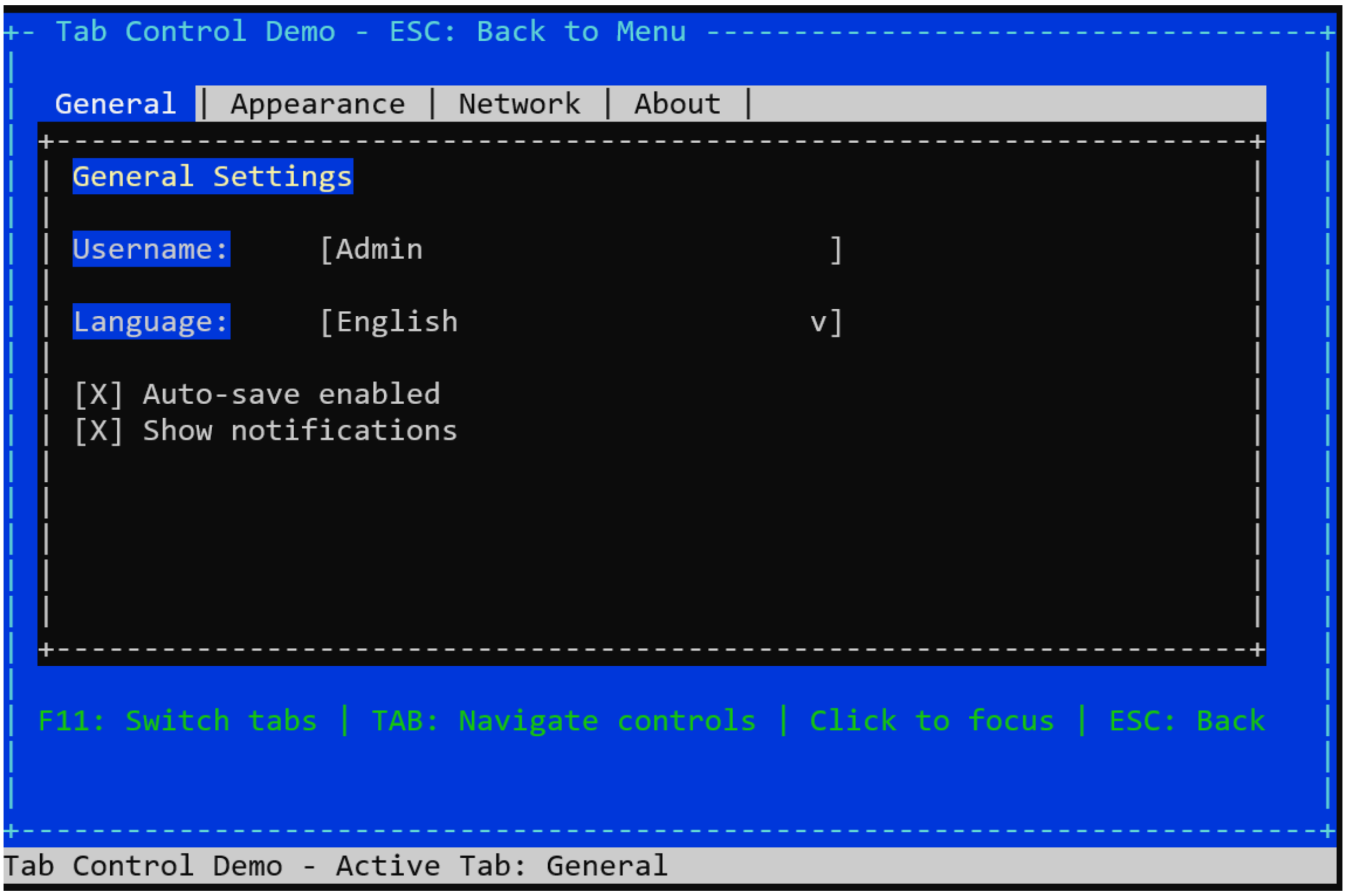

Figure 13. The Tabs Demo.

Figure 14, displays the Controls Demo interface of the Ring TUI Library, showcasing a variety of interactive UI elements rendered in a terminal environment using ASCII graphics. The layout includes three categories of controls: progress bars, spinners, and scroll bars. Progress bars visually represent task completion levels for download, upload, and processing. Spinner controls allow numeric adjustment for quantity and percentage values using [+] and [−] buttons. Horizontal and vertical scroll bars simulate range selection, with current values displayed alongside. Navigation instructions at the bottom indicate support for both keyboard and mouse interaction, including TAB for focus, arrow keys or clicks for adjustment, and ENTER for editing spinner values. This demo highlights the Ring TUI framework's ability to emulate dynamic control widgets in text-based interfaces, enabling responsive and visually informative user interaction within terminal applications. Figure 15 presents the Tabs and the Menu Bar Demo.

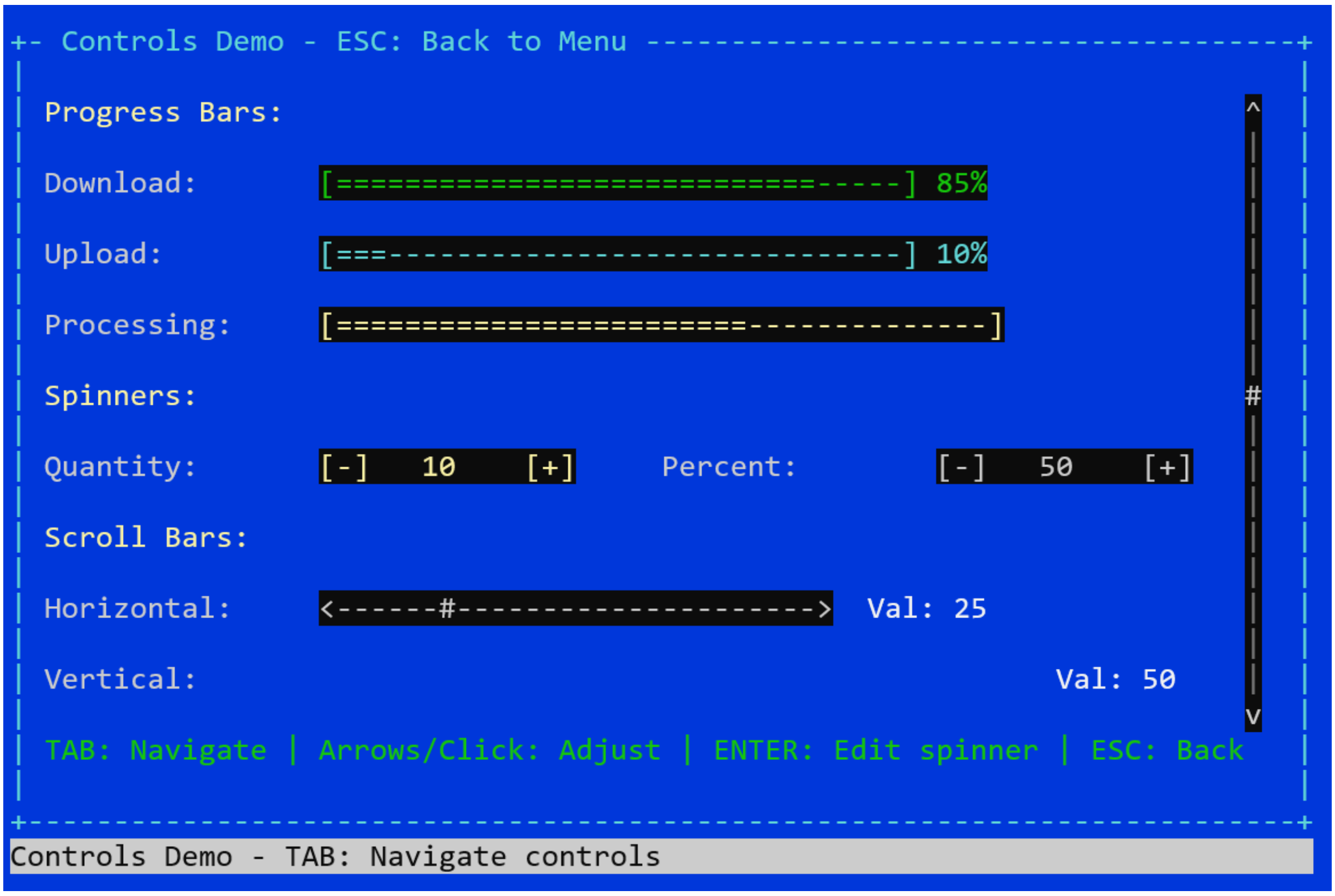

Figure 14. The Controls Demo.

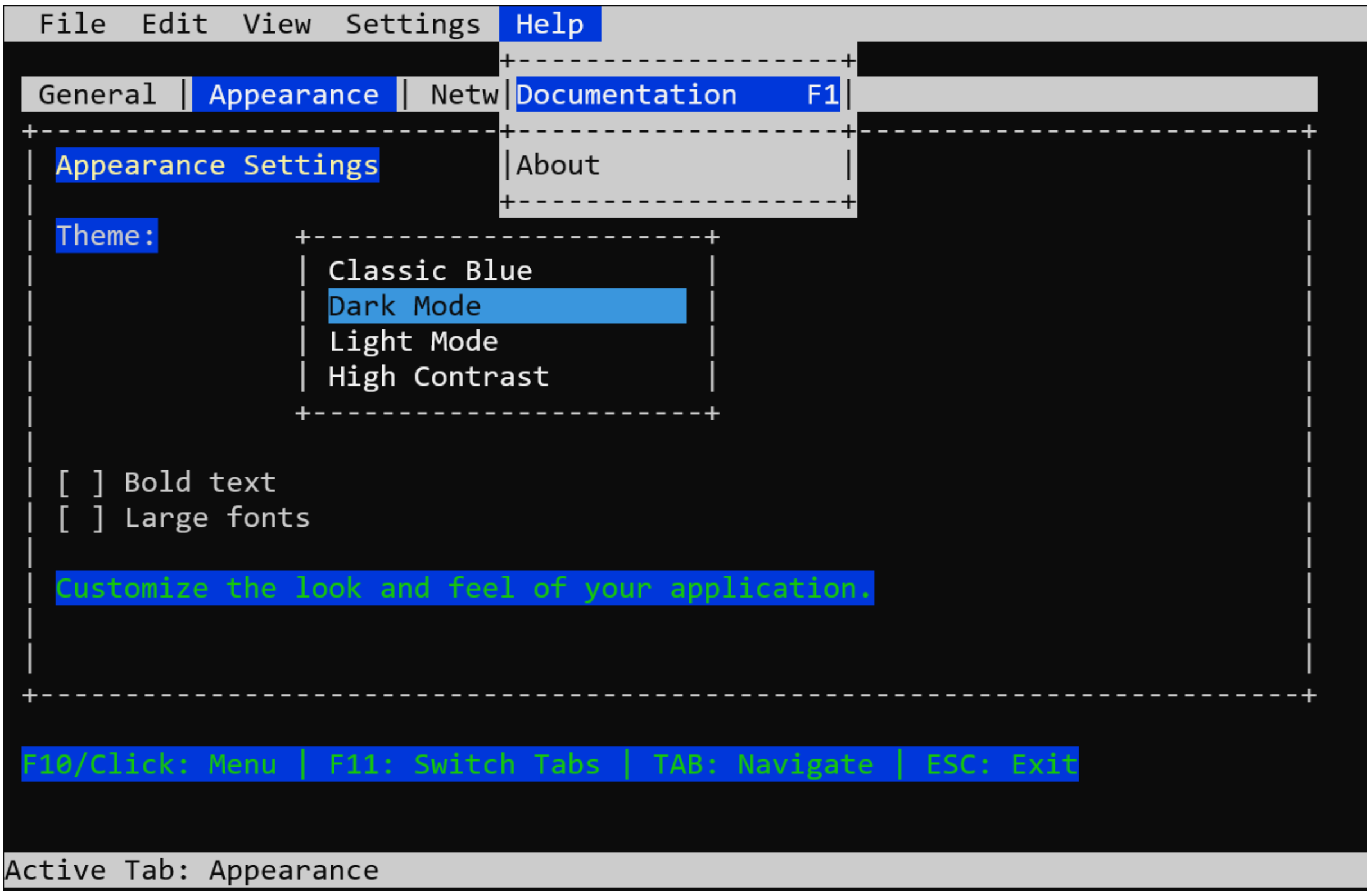

Figure 15. Menu and Tabs Demo.

## 4. Results and Discussion

The development of the Ring TUI framework through a purely prompt-driven workflow provides a rare, detailed view into how a modern large language model behaves when tasked with producing and maintaining a complex, multi-module software system. Across 107 prompts—spanning feature requests, architectural guidance, bug reports, documentation queries, and performance issues—Claude Code (Opus 4.5) demonstrated both impressive generative capabilities and characteristic limitations. The resulting framework, comprising approximately 7,420 lines of Ring code, includes a complete windowing system, event-driven architecture, interactive widgets, data controls, hierarchical menus, a grid/table component, a tree view, tabbed interfaces, and a multi-window desktop environment. This section analyzes the empirical outcomes of the project and discusses their implications for LLM-assisted software engineering.

### 4.1 Quantitative Analysis of the Prompt-Driven Workflow

Table 2 outlines how these prompts align with the project's development phases, highlighting the progression from early bootstrapping to the window-manager-dominated later stages. Table 3 presents an overall statistical summary that integrates all major metrics across the dataset.

Table 2. Development Phases.

| Index | Phase | Description | Count |
|---|---|---|---|
| 1 | Bootstrapping | Kernel, events, basic widgets | 15 |
| 2 | Controls Expansion | Listbox, Combobox, Grid, MenuBar | 20 |
| 3 | Complex UI Systems | Nested menus, TreeView, Tabs | 25 |
| 4 | Window Manager | Dragging, resizing, z-order, redraw | 35 |
| 5 | Final Polish | Focus rules, optimization, performance | 12 |

Table 3. Overall Statistical Summary.

| Index | Metric | Value |
|---|---|---|
| 1 | Total prompts | 107 |
| 2 | Feature prompts | 21 (19.63%) |
| 3 | Bug prompts | 72 (67.29%) |
| 4 | Documentation/info prompts | 9 (8.41%) |
| 5 | Architecture prompts | 4 (3.74%) |
| 6 | Documentation generation prompts | 1 (0.93%) |
| 7 | Largest development phase | Window Manager |
| 8 | Most common bug type | Redraw/flicker |
| 9 | Most common feature type | Widgets/controls |

## 4.2 Capability of LLMs to Produce Cohesive, Multi-Module Systems

One of the most significant findings is that Claude Code was able to generate a fully functional, internally consistent TUI framework without the developer writing a single line of code manually. The model produced:

- A unified event-driven architecture
- A consistent widget hierarchy
- A window manager with dragging, resizing, z-ordering, and taskbar integration
- A menu bar with nested submenus and keyboard shortcuts
- A grid/table component supporting navigation and cell editing
- A tree view with hierarchical expansion and collapse
- A tab control system
- A suite of controls including progress bars, spinners, and scrollbars

The breadth and depth of these components demonstrate that LLMs can handle not only isolated functions but also the architectural coherence required for framework-level development. The model maintained naming conventions, event signatures, and structural patterns across hundreds of generated functions, even when modules were developed days apart. This suggests that LLMs can act as high-productivity junior developers when guided by a human architect who provides constraints, corrections, and long-range direction.

## 4.3 Iterative Refinement as a Core Mechanism

The development process was highly iterative. Approximately 67% of the prompts were bug reports or error-driven corrections. These ranged from simple syntax errors to deep architectural issues such as:

- Infinite redraw loops
- Incorrect event propagation
- Focus inconsistencies
- Mouse hit-testing errors
- Window intersection logic
- Menu navigation state bugs
- Grid editing recursion
- TreeView redraw inefficiencies
- Performance degradation after complex demos

In nearly all cases, Claude Code responded effectively to corrective prompts, producing revised versions that addressed the identified issues. This behavior highlights a key strength of LLM-driven development: the ability to rapidly regenerate or refactor large sections of code in response to high-level feedback. However, the model often repeated certain classes of mistakes—particularly those involving Ring-specific semantics such as:

• The top-down execution model

• The need for Ref() when modifying list items

• The correct naming of RogueUtil constants

These recurring issues underscore the importance of human oversight. While the model can correct errors when pointed out, it does not reliably internalize language-specific rules across long sessions.

## 4.4 Role of the Human Developer: Architect, Not Coder

A clear pattern emerged: the human developer acted as the architect, while the LLM acted as the implementer. The developer provided:

• Architectural constraints (the user should not write loops)

• High-level design decisions (move all event loops into the event manager)

• Domain knowledge (KEY_PGDN is named KEY_PGDOWN in RogueUtil)

• Behavioral expectations (draw each window only once per cycle)

• Performance guidance (avoid redrawing the entire screen unnecessarily)

Claude Code, in turn, translated these constraints into concrete implementations, often producing hundreds of lines of code per prompt. This division of labor mirrors real-world software teams, where senior engineers define system architecture and junior engineers implement modules. The results suggest that LLMs can effectively fill the role of a highly productive junior developer—provided they receive continuous architectural guidance and correction.

## 4.5 Emergent Patterns in LLM Behavior

Analysis of the 107-prompt log reveals several consistent behavioral patterns:

Strengths

• High productivity: The model generated thousands of lines of code in minutes.

• Responsiveness to constraints: It adapted quickly to new architectural rules.

• Long-range coherence: It maintained consistent APIs across modules.

• Refactoring ability: It could restructure entire subsystems on request.

• Conceptual reasoning: It understood event-driven design, widget hierarchies, and UI state machines.

The Weaknesses are:

• Language-specific blind spots: Ring's execution model and reference semantics required repeated correction.

• Tendency toward over-redrawing: Without explicit constraints, the model defaulted to full-screen redraws.

• Occasional architectural drift: Some modules introduced redundant loops or inconsistent naming.

• Performance degradation: As the system grew, the model sometimes introduced inefficient redraw logic.

• Limited memory of earlier constraints: Rules established early in development sometimes had to be restated.

These patterns align with broader observations in LLM research: models excel at generating structure and code volume but require human oversight to maintain architectural integrity and language-specific correctness.

## 4.6 Prompt-Driven Development as a Methodology

This case study demonstrates that prompt-driven development is not only feasible but highly effective for building complex frameworks—provided the workflow is structured around iterative refinement. The key methodological insights include:

• Prompt granularity matters: Small, focused prompts produced more reliable code than broad requests.

• Error-driven iteration is essential: The model improved most rapidly when guided by concrete runtime errors.

• Documentation prompts enhance consistency

• Feature layering works well: Building the framework in layers (widgets → menus → windows → desktop) allowed the model to reuse patterns effectively.

• Human-AI collaboration is synergistic: The human provided architectural vision; the model provided implementation speed.

The success of this workflow suggests that prompt-driven development may become a viable paradigm for accelerating ecosystem growth in emerging programming languages, especially those with limited tooling.

## 4.7 Implications for the Ring Ecosystem

The resulting TUI framework significantly expands the capabilities of the Ring programming language. It provides:

- A full terminal-based UI toolkit
- A window manager suitable for text-mode applications
- A consistent widget library
- A foundation for building IDEs, dashboards, and interactive tools
- A demonstration of how LLMs can accelerate library development for niche languages

This project illustrates how LLMs can serve as force multipliers for language creators and maintainers, enabling rapid development of high-value tooling that would otherwise require substantial engineering effort.

## 4.8 Limitations and Future Directions

While the results are promising, several limitations remain:

- The model required continuous correction for language-specific details.
- Performance optimization still depended heavily on human insight.

Future work may explore:

- Automated testing integrated into the prompt loop
- Multi-agent LLM collaboration (e.g., one model for architecture, one for code)
- Fine-tuning models on Ring-specific corpora
- Extending the framework to support themes, animations, or accessibility features

The development of the Ring TUI framework demonstrates that modern LLMs can successfully generate and maintain large, multi-module software systems when guided through a structured, iterative prompt-driven workflow. Claude Code (Opus 4.5) exhibited strong generative capabilities, effective responsiveness to feedback, and the ability to maintain architectural coherence across thousands of lines of code. The human developer's role shifted from writing code to providing architectural direction, validating outputs, and correcting language-specific issues. This case study provides empirical evidence that LLM-assisted development can meaningfully accelerate the creation of production-grade tooling for emerging programming languages, offering a new model for human-AI collaboration in software engineering.

# 5. Conclusion

This study demonstrates that modern large language models—specifically Claude Code (Opus 4.5)—can generate and maintain complex, multi-module software systems through a purely prompt-driven workflow. By guiding the model with natural-language specifications, architectural constraints, and iterative corrections, we successfully developed a 7,420-line Terminal User Interface (TUI) framework for the Ring programming language without writing any code manually. The resulting system includes a complete windowing environment, an event-driven architecture, a suite of interactive widgets, hierarchical menus, a grid/table component, a tree view, tab controls, and a multi-window desktop interface. These outcomes provide concrete evidence that LLMs can contribute meaningfully to the creation of production-grade tooling for emerging programming languages like the Ring programming language.

The development process revealed a clear division of roles: the human developer acted as the architect—defining structure, identifying errors, and enforcing language-specific rules—while the LLM served as a highly productive implementer capable of generating large volumes of coherent code. This collaborative dynamic proved essential, as the model excelled at producing structure and functionality but required human oversight to maintain architectural integrity and adhere to Ring's execution semantics. The iterative cycle of prompting, testing, error reporting, and refinement emerged as a powerful methodology for steering the model toward correct and maintainable solutions. The findings also highlight both the strengths and limitations of current LLM-based development. Claude Code demonstrated strong reasoning abilities, rapid code generation, and effective adaptation to feedback, enabling the construction of a sophisticated framework in a remarkably short time. At the same time, the model exhibited recurring challenges with language-specific details, performance-sensitive logic, and long-range consistency—areas where human expertise remained indispensable. These observations suggest that LLM-assisted development is most effective when the human provides architectural vision and domain knowledge, while the model handles implementation and refactoring. Beyond the technical achievement, this project illustrates the potential of LLMs to accelerate ecosystem growth for niche or emerging programming languages. For Ring, the new TUI framework expands the language's capabilities, enriches its tooling, and lowers the barrier for building interactive terminal applications. More broadly, the case study offers a blueprint for how language creators, researchers, and open-source communities can leverage LLMs to rapidly prototype libraries, frameworks, and educational materials. As LLMs continue to evolve, their role in software engineering is likely to expand from code generation toward deeper forms of architectural reasoning, automated testing, and long-term project maintenance. Future work may explore multi-agent collaboration, fine-tuning on language-specific corpora, or integrating automated verification into the prompt-driven workflow. Nevertheless, the present study provides strong empirical evidence that human-AI collaboration—when structured around iterative refinement and clear architectural guidance—can produce substantial, coherent, and practically useful software systems.